\documentclass{aastex62}

\usepackage[%
        ]{hyperref}

\hypersetup{
	hypertexnames=true,
	plainpages=false,
	naturalnames=true}
\usepackage{textcomp}
\usepackage{savesym}
\savesymbol{tablenum}
\usepackage{siunitx}
\restoresymbol{SIX}{tablenum}
\usepackage{gensymb}
\usepackage{comment}
\usepackage[utf8]{inputenc}
\usepackage{array}
\usepackage{float}
\usepackage{graphicx}
\usepackage{footmisc}
\usepackage{multirow}
\usepackage{subfigure}
\usepackage{fourier} 
\usepackage{makecell}
\newcommand{\msun}{M$_\odot$}
\newcommand{\rsun}{R$_\odot$}
\newcommand{\rearth}{R$_\oplus$}
\newcommand{\mearth}{M$_\oplus$}
\newcommand{\Teff}{T$_\mathrm{eff}$~}
\newcolumntype{L}{>{\centering\arraybackslash}m{1.2cm}}
\newcolumntype{O}{>{\centering\arraybackslash}m{1.5cm}}
\newcolumntype{M}{>{\centering\arraybackslash}m{1.75cm}}
\newcolumntype{N}{>{\centering\arraybackslash}m{2cm}}

\shorttitle{Atmospheric Modeling of TOI-700~d} 
\shortauthors{Suissa et al.}

\begin{document}
\label{firstpage}

\title{The First Habitable Zone Earth-sized Planet from TESS. III: Climate States and Characterization Prospects for TOI-700~d}

\author[0000-0003-4471-1042]{Gabrielle Suissa}
\affiliation{NASA Goddard Space Flight Center, 8800 Greenbelt Rd, Greenbelt, MD 20771, USA}
\affiliation{NASA Goddard Sellers Exoplanet Environments Collaboration}
\affiliation{Goddard Earth Sciences Technology and Research (GESTAR), Universities Space Research Association, Columbia, Maryland, USA}
\author[0000-0002-7188-1648]{Eric T. Wolf}
\affiliation{NASA Goddard Sellers Exoplanet Environments Collaboration}
\affiliation{Laboratory for Atmospheric and Space Physics, Department of Atmospheric and Oceanic Sciences, University of Colorado Boulder, Boulder, CO 80309, USA}
\affiliation{NASA NExSS Virtual Planetary Laboratory, Box 951580, Seattle, WA 98195}
\author[0000-0002-5893-2471]{Ravi kumar Kopparapu}
\affiliation{NASA Goddard Space Flight Center, 8800 Greenbelt Rd, Greenbelt, MD 20771, USA}
\affiliation{NASA Goddard Sellers Exoplanet Environments Collaboration}
\affiliation{NASA NExSS Virtual Planetary Laboratory, Box 951580, Seattle, WA 98195}
\author[0000-0002-2662-5776]{Geronimo L. Villanueva}
\affiliation{NASA Goddard Space Flight Center, 8800 Greenbelt Rd, Greenbelt, MD 20771, USA}
\affiliation{NASA Goddard Sellers Exoplanet Environments Collaboration}
\author[0000-0002-5967-9631]{Thomas Fauchez}
\affiliation{NASA Goddard Space Flight Center, 8800 Greenbelt Rd, Greenbelt, MD 20771, USA}
\affiliation{NASA Goddard Sellers Exoplanet Environments Collaboration}
\affiliation{Goddard Earth Sciences Technology and Research (GESTAR), Universities Space Research Association, Columbia, Maryland, USA}
\author[0000-0002-8119-3355]{Avi M. Mandell}
\affiliation{NASA Goddard Space Flight Center, 8800 Greenbelt Rd, Greenbelt, MD 20771, USA}
\affiliation{NASA Goddard Sellers Exoplanet Environments Collaboration}
\author[0000-0001-6285-267X]{Giada Arney}
\affiliation{NASA Goddard Space Flight Center, 8800 Greenbelt Rd, Greenbelt, MD 20771, USA}
\affiliation{NASA Goddard Sellers Exoplanet Environments Collaboration}
\affiliation{NASA NExSS Virtual Planetary Laboratory, Box 951580, Seattle, WA 98195}
\author[0000-0002-0388-8004]{Emily A. Gilbert}
\affiliation{NASA Goddard Space Flight Center, 8800 Greenbelt Rd, Greenbelt, MD 20771, USA}
\affiliation{NASA Goddard Sellers Exoplanet Environments Collaboration}
\affiliation{Department of Astronomy and Astrophysics, University of
Chicago, 5640 S. Ellis Ave, Chicago, IL 60637, USA}
\affiliation{The Adler Planetarium, 1300 South Lakeshore Drive, Chicago, IL 60605, USA}
\author[0000-0001-5347-7062]{Joshua E. Schlieder}
\affiliation{NASA Goddard Space Flight Center, 8800 Greenbelt Rd, Greenbelt, MD 20771, USA}
\affiliation{NASA Goddard Sellers Exoplanet Environments Collaboration}
\author[0000-0001-7139-2724]{Thomas Barclay}
\affiliation{NASA Goddard Space Flight Center, 8800 Greenbelt Rd, Greenbelt, MD 20771, USA}
\affiliation{NASA Goddard Sellers Exoplanet Environments Collaboration}
\affiliation{University of Maryland, Baltimore County, 1000 Hilltop Cir, Baltimore, MD 21250, USA}
\author[0000-0003-1309-2904]{Elisa V. Quintana}
\affiliation{NASA Goddard Space Flight Center, 8800 Greenbelt Rd, Greenbelt, MD 20771, USA}
\affiliation{NASA Goddard Sellers Exoplanet Environments Collaboration}
\author[0000-0002-7727-4603]{Eric Lopez}
\affiliation{NASA Goddard Space Flight Center, 8800 Greenbelt Rd, Greenbelt, MD 20771, USA}
\affiliation{NASA Goddard Sellers Exoplanet Environments Collaboration}
\author[0000-0001-8812-0565]{Joseph E. Rodriguez}
\affiliation{Center for Astrophysics $\mid$ Harvard \& Smithsonian, 60 Garden St, Cambridge, MA, 02138, USA}
\author[0000-0001-7246-5438]{Andrew Vanderburg}
\affiliation{Department of Astronomy, The University of Texas at Austin, Austin, TX 78712, USA}
\affiliation{NASA Sagan Fellow}

\correspondingauthor{Gabrielle Suissa}
\email{gabrielle.engelmann-suissa@nasa.gov}

\begin{abstract}

We present self-consistent three-dimensional climate simulations of possible habitable states for the newly discovered Habitable Zone Earth-sized planet, TOI-700~d. We explore a variety of atmospheric compositions, pressures, and rotation states for both ocean-covered and completely desiccated planets in order to assess the planet's potential for habitability. For all 20 of our simulated cases, we use our climate model outputs to synthesize transmission spectra, combined-light spectra, and integrated broadband phase curves. These climatologically-informed observables will help the community assess the technological capabilities necessary for future characterization of this planet -- as well as similar transiting planets discovered in the future -- and will provide a guide for distinguishing possible climate states if one day we do obtain sensitive spectral observations of a habitable planet around a M-star. We find that TOI-700~d is a strong candidate for a habitable world and can potentially maintain temperate surface conditions under a wide variety of atmospheric compositions. Unfortunately, the spectral feature depths from the resulting transmission spectra and the peak flux and variations from our synthesized phase curves for TOI-700~d do not exceed 10~ppm. This will likely prohibit the James Webb Space Telescope (JWST) from characterizing its atmosphere; however, this motivates the community to invest in future instrumentation that perhaps can one day reveal the true nature of TOI-700~d, and to continue to search for similar planets around less distant stars.

\end{abstract}


\section{Introduction}
\label{sec:introduction}

The Transiting Exoplanet Survey Satellite \citep[\textit{TESS},][]{ricker:2015} is ushering the field of exoplanets into a new era with a key objective to detect terrestrial planets around the brightest stars in the solar neighborhood. Terrestrial-sized planets discovered by \textit{TESS} pose an exciting opportunity for the community to perform follow up observations with the James Webb Space Telescope (JWST), ground-based observatories, and future flagship missions to characterize their atmospheres. In particular, rocky planets in the Habitable Zone (HZ) of cool stars (M dwarfs) are our most promising targets to search for biosignatures in the near future, because they are in the region around a star where liquid water could exist on the surface of a planet, and the relatively large planet-to-star size ratio and frequent transits makes planets orbiting M-dwarf stars more amenable for observations. The TOI-700 system, located 31.1~pc from the Sun, is a newly discovered \textit{TESS} system of three planets around an M star with an effective temperature of 3480~K. TOI-700~d, the third planet from the star, has an orbital period of 37 days and receives an incident flux of 0.86~S$_\odot$ \citep{gilbert:2020,rodriguez:2020}, which puts it in the conservative HZ \citep{kopparapu:2013,kopparapu:2016,kopparapu:2017,yang_j:2013}. With 11 sectors of \textit{TESS} observations, \citet[][Part I in this series]{gilbert:2020} reports that TOI-700~d has a derived radius of $1.19\pm0.11$~\rearth and an estimated mass of $1.72^{+1.29}_{-0.63}$~\mearth. With two follow-up transits of TOI-700~d using the \textit{Spitzer} Space Telescope, \citet[][Part II in this series]{rodriguez:2020} derive a radius of $1.144^{+0.062}_{-0.061}$~\rearth and calculate an estimated mass of $2.10^{+0.68}_{-0.65}$~\mearth. These estimated masses have been calculated using the \texttt{Forecaster} mass-radius relation \citep{forecaster}; to date mass measurements of the planet have not been obtained \citep{gilbert:2020,rodriguez:2020}. Observations of other, hotter planets suggest that TOI-700~d is rocky, though there are significant uncertainties in extrapolating these observations to colder, habitable-zone planets like TOI-700~d \citep{rogers:2015,owen&wu:2017}.

As TOI-700~d is in the HZ of an M star, we expect the planet to be tidally locked, and because its eccentricity is close to zero \citep{gilbert:2020,rodriguez:2020}, it is possible that it is in synchronous rotation such that one side of the planet always faces the star \citep{dole:1964,kasting:1993,makarov:2012,barnes:2017}. For slow rotators, this synchronous rotation leads to complicated and unique atmospheric circulation patterns -- the Coriolis effect dictating the circulation weakens, clouds congregate around the substellar point, and heat is transported radially from the dayside of the planet \citep{yang_j:2013, yang_j:2014, yang_j:2019b, kopparapu:2016, way:2016, haqq-misra:2018, wolf:2019, komacek&abbot:2019, delgenio:2019}. These inherently three-dimensional effects can only be captured through the usage of three-dimensional general circulation models (GCMs). In this work, we perform GCM simulations to model different planetary atmospheres suitable for TOI-700~d given its parameters. We explore a grid of potentially habitable climate states, with different axes of comparisons such as varying atmospheric compositions, atmospheric pressures, surface conditions, and orbital phenomena. For each simulation, we produce its corresponding synthesized transmission spectra, combined-light (thermal emission and reflected light) phase-dependent spectra, and integrated broadband phase curve. The modeled climates and associated observables will inform the community on how to interpret future spectral retrievals of TOI-700~d, and what technological sensitivity is required to do so. Additionally, these simulations provide a blueprint for understanding the broad range of possible climate states for habitable planets around early M-stars in general.

The history of the field of exoplanets tells us to expect the unexpected, and this very may well hold true upon future characterizations of TOI-700~d. We recognize that our grid of simulations does not represent an exhaustive treatment of all potential realities for TOI-700~d. Due to the limitations of our current 3-D model design, we do not consider O$_2$ dominated post-runaway atmospheres \citep[e.g.,][]{selsis:2007,wordsworth&pierrehumbert:2014,luger&barnes:2015,tian:2015,lincowski:2018}, nor thick $\sim$100~bar (and higher) modern Venus-like \citep[e.g.,][]{kane:2014,lincowski:2018} or sub-Neptunian atmospheres \citep[e.g.,][]{lopez&fortney:2014,owen&mohanty:2016}. We do not treat prognostic photochemistry in our code nor haze formation, precluding us from considering Titan-analog worlds, complex oxygen dominated chemistry, or other photochemical effects. It is also possible that TOI-700~d has suffered from intense atmospheric escape during the pre-main sequence phase of the star and lacks an atmosphere entirely $-$ although we note that presently TOI-700 has no indications of significant stellar activity, deeming it a quiet star \citep{gilbert:2020}. Here, our focus is simply on potentially habitable climates for TOI-700~d. We limit our study to a set of 20 common habitable atmospheric assumptions that have been widely used by the exoplanet climate modeling community in previous 3-D modeling studies of habitable zone boundaries and of other target objects. In this work, we take a first cut at 3-D modeling potential climates of TOI-700~d, evaluating its potential for habitability, and assessing the requirements for future observation and characterization. 

While no model is without limitations, there are several distinct advantages for using a 3-D climate model to inform our synthetic observables. First, it is well known that clouds pose a significant challenge for characterizing the atmospheres of terrestrial class extrasolar planets. This has been shown by theoretical climate modeling of temperate Earth-sized planets \citep{lincowski:2018,fauchez:2019,suissa:2020,komacek:2020}, as well as by observations of featureless spectra of hot super-Earths \citep{kreidberg:2014}. 3-D climate models allow for the self-consistent and prognostic calculation of cloud fields, based on the confluence of temperature, moisture, and atmospheric circulations. The search for habitable worlds is the search for liquid water. However, given typical lapse rates in planetary atmospheres, any planet with abundant surface liquid water will have water clouds. Accurately accounting for clouds in our models is critical for interpreting observations. Furthermore, 3-D climate models allow for the self-consistent calculation of horizontal heat transport, which is important on tidally locked planets, and critical for determining phase-dependent observations. The inclusion of prognostic clouds and horizontal heat transport are necessary for determining day-to-night differences in the thermal energy emitted from these worlds, and for calculating full thermal emission phase curves \citep{selsis:2011, yang_j:2014, koll:2015, turbet:2016, boutle:2017, wolf:2017, wolf:2019}. 

This paper is outlined as follows. In Section~\ref{sec:methods}, we detail the suite of GCM simulations we ran and the radiative transfer tools we used. In Section~\ref{sec:results} we present transmission spectra and combined phase curves for a selection of our models. In Section~\ref{sec:discussion} we discuss the implications of the observability of the atmosphere of TOI-700~d. For more details about the discovery of the TOI-700 system and the Spitzer confirmation of TOI-700~d, please see our companion papers Parts I \& II: \citet{gilbert:2020} \& \citet{rodriguez:2020}.

\section{Methods}
\label{sec:methods}

\subsection{Overview of Climate Models}
\label{sec:gcm}

In this work we use a three-dimensional general circulation and climate system model to simulate self-consistent atmospheres for the newly discovered planet TOI-700~d. We use the ExoCAM\footnote{\url{https://github.com/storyofthewolf/ExoCAM}}$^,$\footnote{\url{https://github.com/storyofthewolf/ExoRT}} modeling package, which is a modified version of the Community Atmosphere Model (CAM) version 4 from the National Center for Atmospheric Research in Boulder, CO \citep{CAM}. ExoCAM includes model configurations, source code modifications, and initial conditions files needed to facilitate exoplanet studies. It has been used in numerous climate studies of terrestrial extrasolar planets \citep{kopparapu:2017,wolf:2017,wolf:2019, haqq-misra:2018, komacek:2019, komacek&abbot:2019, adams:2019, yang_h:2019, kang:2019}. Here, we use ExoCAM to conduct simulations of TOI-700~d considering a variety of atmospheric compositions and planet archetypes that are typically associated with habitable worlds. We consider both aquaplanets (completely ocean-covered with no land) and land-covered planets (completely land-covered with no ocean), with atmospheres composed of varying mixtures of N$_2$, CO$_2$, CH$_4$, H$_2$, and H$_2$O. 

For aquaplanet simulations, we assume a 50~meter deep slab ocean with no ocean heat transport (OHT). Note that OHT can modulate surface temperatures \citep{hu&yang_j:2014, checlair:2019, delgenio:2019, yang_j:2019}, particularly for cooler planets that lack emergent continents. However, we do not expect OHT to significantly affect the resulting transmission spectra. While OHT may indeed bring more clouds toward the terminator region, a significant change in their average altitude is not expected. Yet it is the altitude of the cloud deck that impacts the transmission spectra the most; it has been shown \citep{fauchez:2019} that even a small cloud coverage ($20\%$ or less) is enough to raise the continuum level above the cloud deck. OHT could also affect thermal emission phase curve morphologies \citep[e.g.,][]{yang_j:2019}. However, \citet{fujii:2017} found that stratospheric water vapor on tidally locked planets is not strongly affected by OHT, while \citet{yang_j:2019} determine that the role of OHT diminishes on warm (T$_S$ $>$ 290~K) tidally locked worlds. The presence and location of any emergent continents remains a significant uncertainty for any dynamic ocean model calculations, as continents dramatically change or even halt day-to-night side OHT on tidally locked planets, depending upon the specific continental configuration that is assumed \citep{delgenio:2019,yang_j:2019}. For aquaplanet simulations, we assume that sea ice forms whenever sea surface temperatures fall below the freezing point of seawater at 271.36~K. We also assume the default snow and ice albedo parameterizations, which divide ocean, sea ice, and snow albedos into two bands: visible and near-IR divided at $\sim$\SI{0.7}{\micro\metre}. Following \citet{joshi&haberle:2012,shields:2013,vonparis:2013}, we set the visible (near-IR) sea ice albedo to 0.67 (0.3), the snow albedo to 0.8 (0.68), and the ocean albedo to 0.07 (0.06). For land planet simulations we assume a completely desiccated planet, with a uniform surface composed of sand with an albedo of 0.3 at all wavelengths. 

The ExoCAM radiative transfer module, originally constructed for early Earth studies \citep{wolf&toon:2013}, uses 28 spectral intervals across all wavelengths, with HITRAN 2004 absorption coefficients for H$_2$O, CO$_2$, and CH$_4$. Water vapor continua are included using the MT$_\_$CKD model version 2.5 \citep{clough:2005}. H$_2$O is assumed to be pressure broadened by Earth-air. CO$_2$ self-broadening is taken to be 1.3 times the foreign-broadening component while H$_2$O foreign broadening is assumed to be that of Earth-air \citep{Kasting:1984}. We follow the “MTCKD” parameterization described in \citet{halevy:2009}, and at high-CO$_2$ our method produces intermediate-strength absorption compared to other popular CO$_2$ continuum parameterizations \citep{halevy:2009}. Collision induced absorption are included for N$_2$-N$_2$, H$_2$-H$_2$, and N$_2$-H$_2$ pairs \citep{borysow:1986a, borysow:1986b}. Note that while this particular set of absorption coefficients is older and biases towards overestimating near-IR absorption in water-rich atmospheres \citep{yang_j:2016}, model comparisons of full 3-D climate simulations against other models with updated absorption coefficients are in reasonable agreement for planets around M-dwarf stars, including for high-CO$_2$ worlds \citep{wolf:2018, fauchez_thai:2019b}.

Simulations were run with 4\degree$\times$5\degree ~horizontal resolution and 51 vertical layers extending from the surface up to $\sim$0.01~mbar pressures, using a finite-volume dynamical core \citep{lin&rood:1996}. The high model top allows for a better representation of upper level water vapor and clouds, which is critical for calculating synthetic transmission spectra \citep{suissa:2020}. All simulations are initiated from present-day Earth global mean surface temperatures, and run until the top-of-atmosphere energy balance reaches equilibrium, taking 30 to 100 years depending on the specific simulation. Note that recent works indicate that Snowball hysteresis does not occur for tidally locked planets around M dwarfs \citep{checlair:2017, checlair:2019}; thus we expect our results to be unchanged for different assumptions of the initial temperature fields. The atmospheric constituents N$_2$, CO$_2$, CH$_4$, and H$_2$ are assumed to be well mixed in the atmosphere. However, H$_2$O and water clouds vary self-consistently with the ambient meteorological conditions calculated in the model. Liquid cloud particle sizes are set to a constant value of \SI{14}{\micro\metre}, while ice cloud particle sizes rely on a temperature-dependent function and can range in size from up to several hundred \SI{}{\micro\metre}. 

Table~\ref{tab:system_properties} lists the stellar and planetary system properties used in our 3-D simulations. These parameters vary slightly compared to those presented in the final versions of the discovery paper, \citet{gilbert:2020}, and the \textit{Spitzer} confirmation paper, \citet{rodriguez:2020}, as they have continued to refine their estimates. Still, our values remain within 1$\sigma$ of the final numbers. We note that the additional data and independent analysis performed by \citet{rodriguez:2020} yields parameters that are largely consistent with those of \citet{gilbert:2020} and of this work (see Table~\ref{tab:system_properties}). We remind the reader that the planetary mass is not a derived or fitted parameter; it has been calculated by both \citet{gilbert:2020} and \citet{rodriguez:2020} using the \texttt{Forecaster} mass-radius relation \citep{forecaster}. For the input stellar spectra we interpolate from the BT-Settl spectral models \citep{allard:2007} to construct a TOI-700 specific spectrum, with \Teff = 3480~K, [Fe/H] = 0, and log g = 4.81. We assume that TOI-700~d is synchronously rotating, meaning the planetary rotation period equals its orbital period (37.43 Earth days), for all cases except where specifically noted. We also run two simulations where we assume that the planet is in a 2:1 spin-orbit resonance, meaning that the planet experiences one diurnal period per orbit. 

\begin{table}
\centering
\setlength{\extrarowheight}{5pt}
\begin{tabular}{|c|c|c|}
\hline
\textit{Stellar Parameters} & This work & Rodriguez et al. (2020) \\
\hline
Mass (M$_\odot$) & 0.4151 & $0.415\pm 0.020$ \\
Radius (R$_\odot$) & 0.4185 & $0.424\pm 0.017$ \\
Luminosity (L$_\odot$) & 0.0232 & $0.0232^{+0.0027}_{-0.0025}$ \\
log \textit{g} & 4.81 & $4.802^{+0.038}_{-0.039}$ \\ 
Metallicity (dex) & 0 & $-0.07\pm 0.11$ \\ 
\hline
\hline
\textit{Planetary Parameters} & This work & Rodriguez et al. (2020) \\
\hline
Period (days) & 37.43 & $37.42475^{+0.00036}_{-0.00040}$ \\
Mass (M$_\oplus$) & 1.423 & $1.72^{+1.29}_{-0.63}$ \\ 
Radius (R$_\oplus$) & 1.115 & $1.94^{+0.69}_{-0.57}$ \\
Surface gravity (m/s$^2$) & 11.24 & -- \\
Semi major axis (AU) & 0.163356 & $0.163^{+0.0026}_{-0.0027}$ \\
Stellar flux (W/m$^2$) & 1183.36 & $1160^{+140}_{-130}$\\
\hline
\end{tabular}
\caption{Parameters for TOI-700 and TOI-700~d used in this work's simulations. These parameters vary compared to those presented in the final versions of \citet{gilbert:2020} and \citet{rodriguez:2020}, yet they remain within 1$\sigma$ of the published values. In the rightmost column, we display the parameters from \citet{rodriguez:2020}, which were derived using the \textit{Spitzer} light curve in combination with the \textit{TESS} observations.}
\label{tab:system_properties}
\end{table}

Our simulation grid is summarized in Table~\ref{tab:grid}. We consider three different archetypes of atmospheric compositions. ``Modern Earth'' in this work refers to an atmosphere with a N$_2$ dominated pressure of 1~bar, 400~ppm CO$_2$, 1.7~ppm CH$_4$. O$_2$ is not considered in this study because it does not significantly affect planetary climate, acting primarily as background gas contributing to pressure broadening and molecular scattering effects. ``Archean Earth'' in this work is inspired by our understanding of our planet's early history, where higher levels of CO$_2$ and CH$_4$ than we experience today were required to keep the planet warm despite the faint young Sun \citep{Charnay:2013, wolf&toon:2013}. ``Early Mars'' is loosely used to describe a CO$_2$ dominated atmosphere, which would have been required for liquid water to exist on early Mars \citep{pollack:1987,halevy:2009,forget:2013,wordsworth:2013,wordsworth:2016,haberle:2017}. The three above archetypes are explored both as completely ocean-covered aquaplanets and as completely desiccated land-covered planets. We also include a ``plain'' aquaplanet case, with only H$_2$O and N$_2$ and an atmospheric pressure of 1~bar, and an ``H$_2$-supporting'' case with atmospheric abundances similar to that of an ``Archean Earth'' except with 0.1~bar of H$_2$, inspired by solutions for the faint young Sun paradox posed by \citet{wordsworth&pierrehumbert:2013}. In addition, we examine the ``Archean Earth'' and ``Early Mars'' archetypes with different atmospheric pressures for both desiccated and aquaplanet planets, ranging from 0.5~bar to 10~bar in some cases. Finally, we analyze the effect of orbital phenomenon for one of the aqua and land ``Archean Earth'' cases, by experimenting with a 2:1 rotation:orbital resonance instead of synchronous rotation. 

Our GCM simulations include N$_2$ and Earth's most significant greenhouse gases: CO$_2$, CH$_4$ and H$_2$O. We do not include other gases such as O$_2$, O$_3$, or C$_2$H$_6$ because these gases are not expected to impact the climate as strongly and/or because their abundances are challenging to predict in a self-consistent way without a photochemical model. We anticipate that including photochemistry would affect the abundances and vertical profiles of gases in our atmospheres, and this is an important avenue of future work. For example, \citet{Segura:2005} show that a modern-Earth-like planet orbiting an M dwarf can result in the planet's atmosphere accumulating over two orders of magnitude more CH$_4$ compared to the same planet around the Sun. This is because M dwarfs produce less radiation near \SI{300}{\nano\metre} where O$_3$ is photolyzed. Oxygen radicals generated by this photolysis are the dominant sink of CH$_4$ in modern Earth's atmosphere; with the decrease of this photolysis rate comes the potential increase in CH$_4$ abundance for planets around M dwarfs. Additionally, organic haze formation can initiate if the CH$_4$/CO$_2$ ratio exceeds 0.1 \citep{trainer:2004,Arney:2016,Arney:2017}; however in our simulated atmospheres, methane abundances are kept below this level, and thus photochemical hazes would not be expected to form. 

\begin{table}
\centering
\setlength{\extrarowheight}{5pt}
\begin{tabular}{|*{7}{c|}}
\hline
 & \multicolumn{3}{c|}{\textbf{Aqua}} & \multicolumn{3}{c|}{\textbf{Desiccated}} \\
  & \textit{Atmospheric Specifications} & \multicolumn{2}{c|}{\textit{Atm. Pressure}}  & \textit{Atmospheric Specifications} & \multicolumn{2}{c|}{\textit{Atm. Pressure}}  \\
\hline
\hline
``Modern Earth'' & 400~ppm CO$_2$, 1.7~ppm CH$_4$ & \multicolumn{2}{c|}{1~bar} & 400~ppm CO$_2$, 1.7~ppm CH$_4$ & \multicolumn{2}{c|}{1~bar} \\
\hline
\hline
\multirow{7}{*}{``Archean Earth''} & 0.01~bar CO$_2$, 0.0001~bar CH$_4$ & \multicolumn{2}{c|}{1~bar} & \multirow{7}{*}{0.1~bar CO$_2$, 0.001~bar CH$_4$} & \multirow{7}{*}{1~bar} & \multirow{3}{*}{Synch.}  \\
\cline{2-4}
 & \multirow{5}{*}{0.1~bar CO$_2$, 0.001~bar CH$_4$} & \multicolumn{2}{c|}{0.5~bar} & & & \\
 & & \multirow{2}{*}{1~bar} & Synch. & & & \\ 
 \cline{4-4}
  \cline{7-7}
 & & & 2:1 & & & \multirow{4}{*}{2:1} \\
 & & \multicolumn{2}{c|}{4~bar} & & &\\
 & & \multicolumn{2}{c|}{10~bar} & & &\\ 
\cline{2-4}
 & 0.1~bar CO$_2$, 0.002~bar CH$_4$ & \multicolumn{2}{c|}{1~bar} & & &\\
\hline
\hline
\multirow{4}{*}{``Early Mars''} & \multirow{4}{*}{CO$_2$ dominated} & \multicolumn{2}{c|}{0.5~bar} & \multirow{4}{*}{CO$_2$ dominated} & \multicolumn{2}{c|}{1~bar}\\
& & \multicolumn{2}{c|}{1~bar} & & \multicolumn{2}{c|}{4~bar}\\
& & \multicolumn{2}{c|}{2~bar} && \multicolumn{2}{c|}{10~bar} \\
& & \multicolumn{2}{c|}{4~bar} && \multicolumn{2}{c|}{} \\
\hline
\hline
``H$_2$-supporting'' & \vtop{\hbox{\strut ~~~~~~~~~~~~~~~~~~~~~~~N$_2$ dominated }\hbox{\strut 0.1~bar CO$_2$, 0.001~bar CH$_4$, 0.1~bar H$_2$}} & \multicolumn{2}{c|}{1~bar} & \multicolumn{3}{c|}{} \\
\hline
\hline
``Plain'' & N$_2$ dominated & \multicolumn{2}{c|}{1~bar} & \multicolumn{3}{c|}{} \\
\hline
\end{tabular}
\caption{Grid of simulations used in this work. We explored different compositions, surface types (aquaplanet vs desiccated), atmospheric pressures, and orbital phenomenon (synchronously rotating or in a 2:1 rotation:orbital resonance).}
\label{tab:grid}
\end{table}

\subsection{The Planetary Spectrum Generator and GlobES}
\label{sec:psg}
We synthesize transmission and emission spectra with the Planetary Spectrum Generator \citep[PSG, \url{https://psg.gsfc.nasa.gov/}, ][]{psg:2018}. PSG is a spectroscopic suite that integrates the latest radiative transfer methods and spectroscopic parameterizations, and includes a realistic treatment of multiple scattering in layer-by-layer spherical geometry. Specifically, multiple scattering from atmospheric aerosols is enabled within PSG using the discrete ordinates method, in which the radiation field is approximated by a discrete number of streams distributed in angle with respect to the plane-parallel normal. The angular dependence of the scattering phase function for a particular aerosol is described in terms of an expansion in terms of Legendre Polynomials, typically with the number of expansion terms equal to the number of stream pairs. As implemented in PSG, the Legendre expansion coefficients are pre-computed using an assumed particle size distribution for each available aerosol type \citep[e.g.,][]{massie:2013}. PSG permits the ingestion of billions of spectral lines of over 1,000 species from several spectroscopic repositories (e.g., HITRAN, JPL, CDMS, GSFC-Fluor). For this investigation, the molecular spectroscopy is based on the latest HITRAN database \citep{gordon:2017}, which is complemented by UV/optical data from the MPI database \citep{keller-rudek:2013}.

In order to capture the heterogeneous properties of the atmosphere and the surface as determined by the GCM, the GlobES (Global Exoplanetary Spectra, \url{https://psg.gsfc.nasa.gov/apps/globes.php}) module of PSG was used to compute combined emission and reflection spectra across the observable disk. PSG currently provides templates and conversion scripts for several GCM models, including ROCKE-3D \citep{ROCKE3D:2017}, Laboratoire de Météorologie Dynamique \citep[LMD,][]{forget:1999,hourdin:2006,wordsworth:2011}, and the Community Atmosphere Model \citep[CAM,][]{CAM} as used for this investigation. We ingested the ExoCAM GCM-generated vertical profiles of temperature, pressure, volume mixing ratios of molecular species, and mass mixing ratios and particles sizes for liquid and water ice clouds into GlobES at every 4$\degree$ of latitude across the terminator to create a local spectrum. We then averaged all of the resulting local spectra to create an average terminator spectrum for the planet. For direct imaging and secondary eclipse simulations, GlobES performs radiative transfer simulations across the whole observable disk considering the appropriate incidence, emission, and phase angles for that grid point, while the individual spectra are integrated considering the projected area of each bin. 

When creating combined emitted and reflected light spectra through GlobES, we calculated the contrast in radiance of the planet at every 5\degree~of its phase in its orbit around the star. We implemented a two-stream approximation to model the angular dependence of aerosol scattering with two Legendre polynomials to approximate the phase function. For each phase's emission and reflected spectrum, we then took the average of the contrast values between \SI{15}{\micro\metre} and \SI{21}{\micro\metre}. We chose this wavelength band because thermal emission is more receptive at the mid-infrared, and because it corresponds to JWST's Mid-Infrared Instrument (MIRI) F1800W imaging filter, which has an average photo conversion efficiency (PCE) of $\sim$0.31 \citep{bouchet:2015}. For clarity purposes, the phase curves are shown not to drop to zero at secondary eclipse (phase of 0), even though the planet would be behind the star. Note that in the phase curves presented in this work, transit occurs at the phase of 180, contrary to the phase-related definitions that other works may use \citep[e.g.,][]{kreidberg:2019}.

\section{Results}
\label{sec:results}
\subsection{Climate States}
\label{sec:climate}

\begin{figure}
    \centering
    \resizebox{0.8\linewidth}{!}{\includegraphics{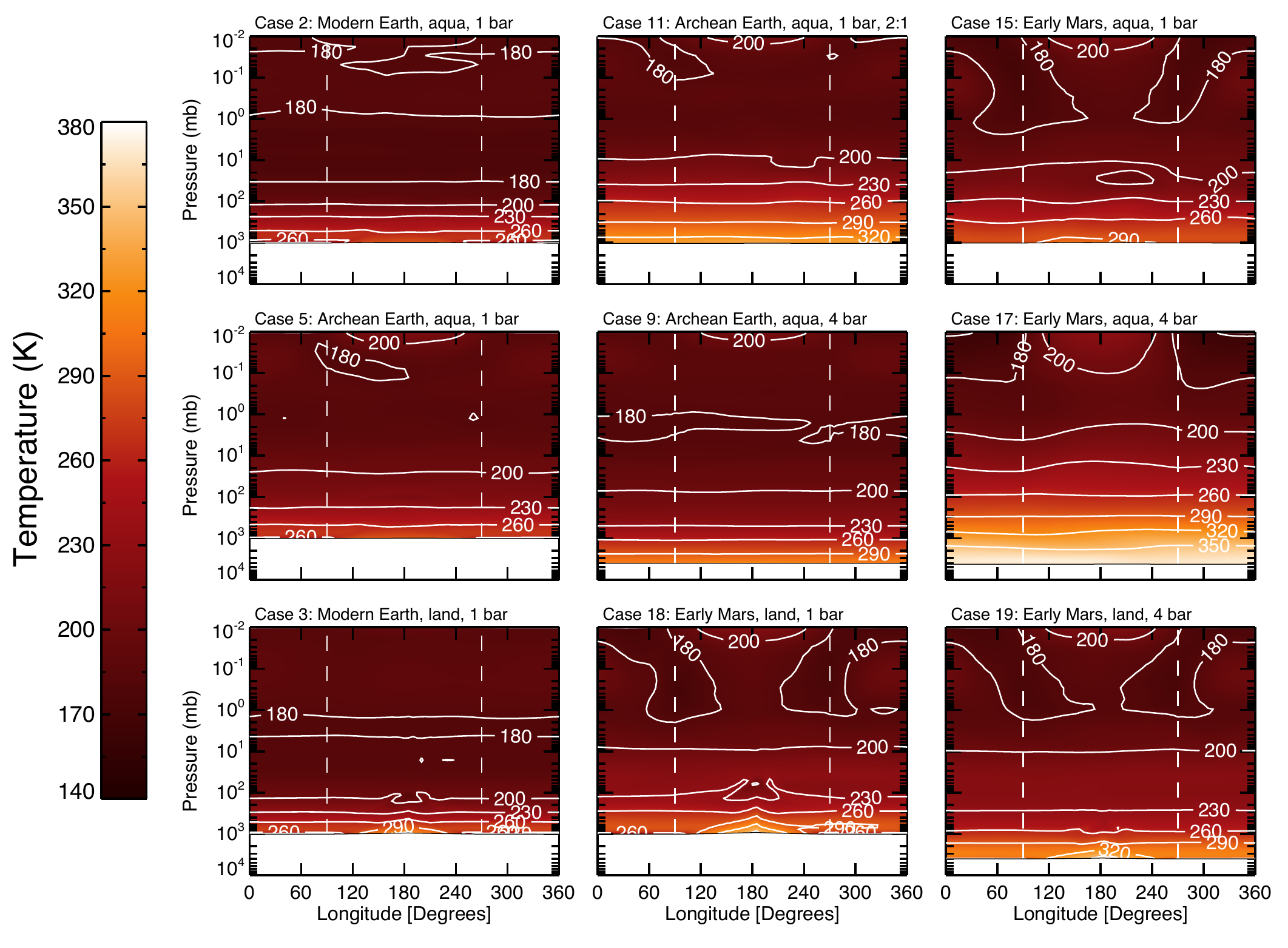}}
    \caption{Vertical temperature profiles taken along the equator from 9 characteristic cases. The substellar point is located at 180 degrees. The vertical dashed lines mark the terminators. Despite varying surface pressures and temperatures, stratospheric temperatures are remarkably similar, hovering at $\sim$180~K.}
    \label{fig:ts}
\end{figure}

\begin{figure}
    \centering
    \resizebox{0.8\linewidth}{!}{\includegraphics{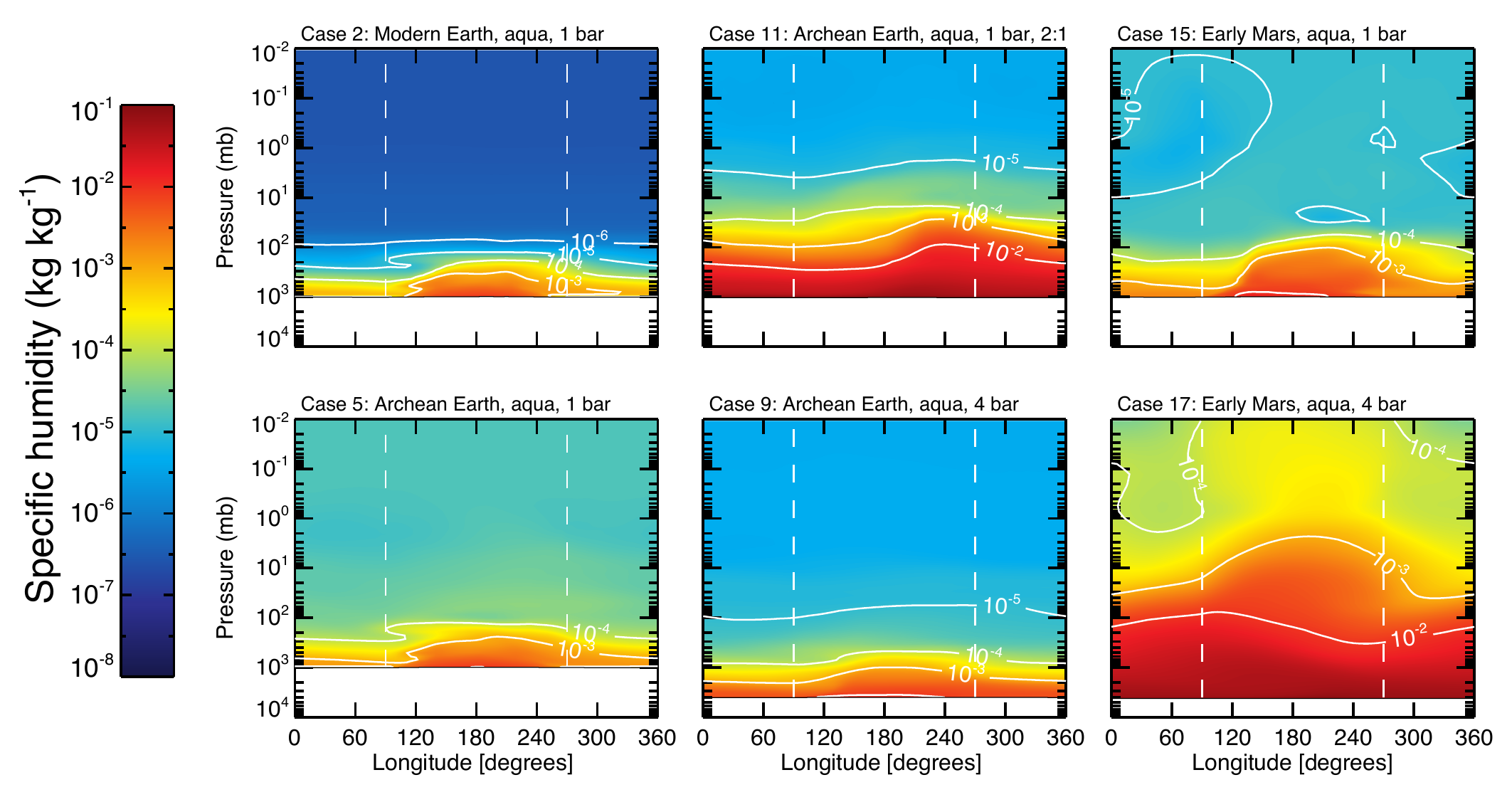}}
    \caption{Vertical profiles of water vapor taken along the equator. Same as Figure~\ref{fig:ts} except dry cases have been omitted. The substellar point is located at 180 degrees. The vertical dashed lines mark the terminators.}
    \label{fig:q}
\end{figure}

\begin{figure}
    \centering
    \resizebox{0.8\linewidth}{!}{\includegraphics{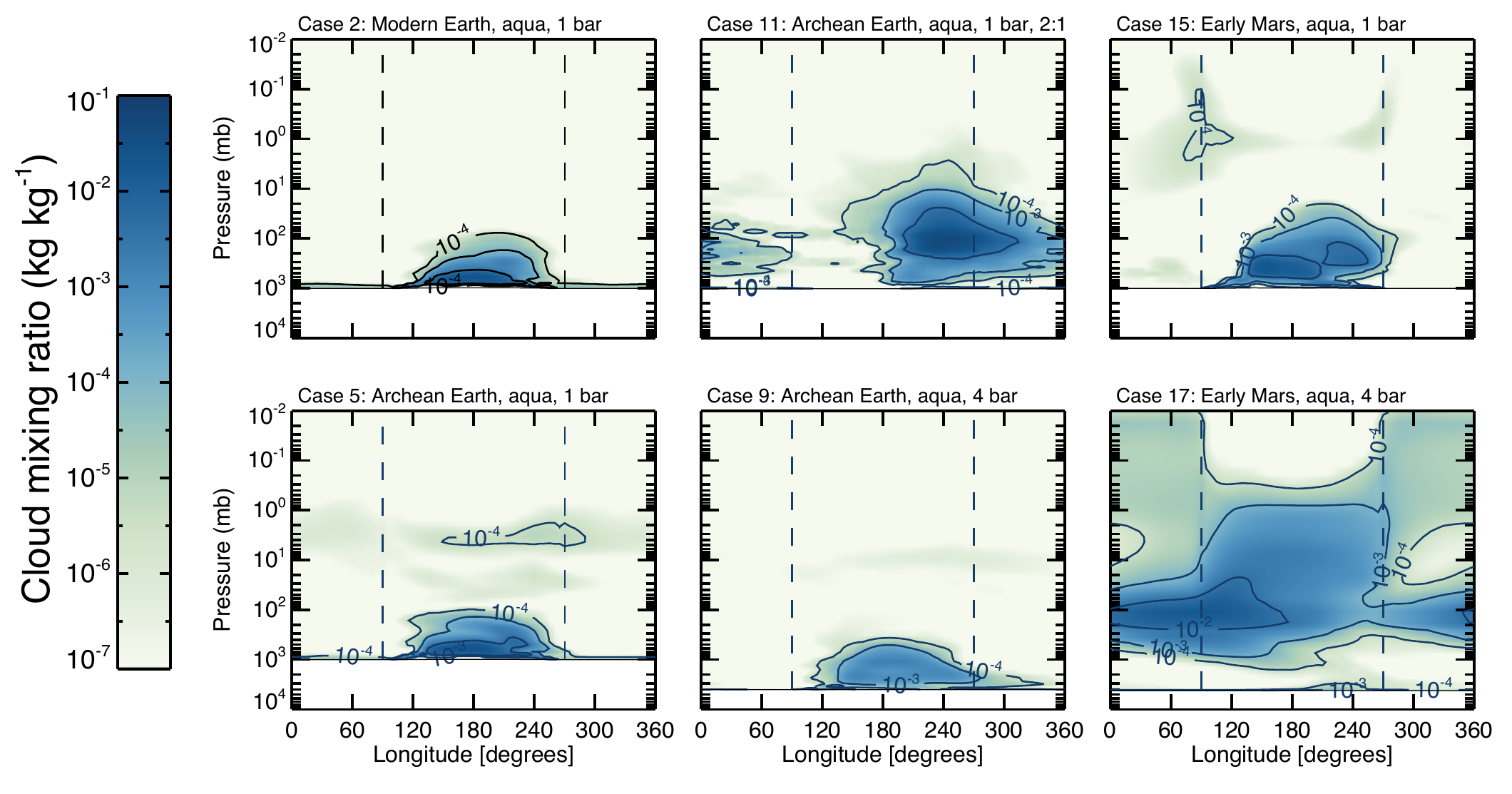}}
    \caption{Vertical profiles of cloud water taken along the equator. Same as Figure~\ref{fig:ts} except dry cases have been omitted. The substellar point is located at 180 degrees. The vertical dashed lines mark the terminators.}
    \label{fig:cld}
\end{figure}

In this section we briefly summarize our 3-D climate modeling results. In order to understand our synthetic observables, we must first understand the underlying atmospheric conditions. In Table~\ref{tab:climate_properties}, we describe the basic configuration and climatological properties from our 20 simulations. In Figures~\ref{fig:ts},~\ref{fig:q}, and~\ref{fig:cld}, we show vertical profiles of temperature, water vapor, and cloud condensate taken in a slice along the equator for selected simulations. TOI-700~d is a strong candidate for a habitable world, and can potentially maintain temperate surface conditions and significant fractions of surface liquid water under a wide variety of atmospheric compositions. This should come as little surprise, because TOI-700~d is located well within the Habitable Zone \citep{kopparapu:2013, kopparapu:2016, yang_j:2013} of its parent star.

\begin{table}
\centering
\begin{tabular}{|c|O|L|L|M|M|c|M|L|L|M|}
\hline
\# & Archetype & Surface type & Surface pressure (bar) & Primary atmospheric constituent & ~~~Minor \newline constituents, partial pressures (bar) & Rotation & Surface temperature (K) & TOA albedo & Sea ice fraction (\%) & Stratospheric water vapor (kg/kg) \\
\hline
1 & Plain & aqua & 1 & N$_2$ & H$_2$O & synchronous & 236.7 & 0.3968 & 76.2\%& $6.5014\mathrm{e}-8$ \\
\hline
2 & Modern Earth & aqua & 1 & N$_2$ & ~~~~H$_2$O \newline ~CO$_2$ (4e-4) \newline CH$_4$ (1.7e-6) & synchronous & 246.7 & 0.3893 & 71.2\% & $3.605\mathrm{e}-7$ \\ 
\hline
3 & Modern Earth & land & 1 & N$_2$ & ~~~~H$_2$O \newline ~CO$_2$ (4e-4) \newline CH$_4$ (1.7e-6) & synchronous & 232.7 & 0.2840 & N/A & N/A \\
\hline
4 & Archean Earth & aqua & 1 & N$_2$ & ~~~~H$_2$O \newline ~CO$_2$ (0.01) \newline CH$_4$ (1.0e-4) & synchronous & 258.4 & 0.3594 & 66.3\% & $7.316\mathrm{e}-7$ \\
\hline
5 & Archean Earth & aqua & 1 & N$_2$ & ~~~~H$_2$O \newline ~CO$_2$ (0.1) \newline CH$_4$ (1.0e-3) & synchronous & 263.2 & 0.3988 & 62.3\% & $2.644\mathrm{e}-5$ \\
\hline
6 & Archean Earth & aqua & 1 & N$_2$ & ~~~~H$_2$O \newline ~CO$_2$ (0.1) \newline CH$_4$ (2.0e-3) & synchronous & 263.7 & 0.3998 & 61.6\% & $3.645\mathrm{e}-5$ \\ 
\hline
7 & Archean Earth & land & 1 & N$_2$ & ~~~~H$_2$O \newline ~CO$_2$ (0.1) \newline CH$_4$ (1.0e-3) & synchronous & 251.5 & 0.2455 & N/A & N/A \\
\hline
8 & Archean Earth & aqua & 0.5 & N$_2$ & ~~~~H$_2$O \newline ~CO$_2$ (0.1) \newline CH$_4$ (1.0e-3) & synchronous & 256.8 & 0.3906 & 67.3\% & $5.615\mathrm{e}-5$ \\
\hline
9 & Archean Earth & aqua & 4 & N$_2$ & ~~~~H$_2$O \newline ~CO$_2$ (0.1) \newline CH$_4$ (1.0e-3) & synchronous & 307.2 & 0.317 & 0 & $6.406\mathrm{e}-6$ \\
\hline
10 & Archean Earth & aqua & 10 & N$_2$ & ~~~~H$_2$O \newline ~CO$_2$ (0.1) \newline CH$_4$ (1.0e-3) & synchronous & 360.8 & 0.2452 & 0 & $2.203\mathrm{e}-6$ \\
\hline
11 & Archean Earth & aqua & 1 & N$_2$ & ~~~~H$_2$O \newline ~CO$_2$ (0.1) \newline CH$_4$ (1.0e-3) & 2:1 resonant & 325.7 & 0.131 & 0 & $6.095\mathrm{e}-6$ \\ 
\hline
12 & Archean Earth & land & 1 & N$_2$ & ~~~~H$_2$O \newline ~CO$_2$ (0.1) \newline CH$_4$ (1.0e-3) & 2:1 resonant & 257.0 & 0.2547 & N/A & N/A \\ 
\hline
13 & H$_2$-supporting & aqua & 1 & N$_2$ & ~~~~H$_2$O \newline ~CO$_2$ (0.1) \newline CH$_4$ (1.0e-3) \newline H$_2$ (0.1) & synchronous & 267.6 & 0.394 & 60.3\% & $2.669\mathrm{e}-5$ \\ 
\hline
14 & Early Mars & aqua & 0.5 & CO$_2$ & H$_2$O & synchronous & 266.2 & 0.3572 & 62.6\% & $2.802\mathrm{e}-5$ \\ 
\hline
15 & Early Mars & aqua & 1 & CO$_2$ & H$_2$O & synchronous & 284.4 & 0.3775 & 0.05\% & $2.520\mathrm{e}-5$ \\ 
\hline
16 & Early Mars & aqua & 2 & CO$_2$ & H$_2$O & synchronous & 324.3 & 0.2771 & 0\% & $3.027\mathrm{e}-5$ \\
\hline
17 & Early Mars & aqua & 4 & CO$_2$ & H$_2$O & synchronous & 364.2 & 0.2214 & 0\% & $2.857\mathrm{e}-4$ \\ 
\hline
18 & Early Mars & land & 1 & CO$_2$ & none & synchronous & 258.9 & 0.2304 & N/A & N/A \\
\hline
19 & Early Mars & land & 4 & CO$_2$ & none & synchronous & 302.3 & 0.2165 & N/A & N/A \\
\hline
20 & Early Mars & land & 10 & CO$_2$ & none & synchronous & 353.5 & 0.2310 & N/A &N/A \\
\hline
\end{tabular}
\caption{Global mean climatological properties for each of our simulations. By definition, sea ice fraction and stratospheric water vapor or not applicable for desiccated land planet cases. The stratospheric water vapor is taken at the model top, 0.01~mbar. All quantities given are the global and temporal mean.}
\vspace{-2pt}
\label{tab:climate_properties}
\end{table}

Our 14 aquaplanet cases have global mean surface temperatures that range from 236.7~K to 364.2~K, but all still are technically habitable worlds. In our coldest simulation, 1~bar N$_2$ ``plain'' aquaplanet with no CO$_2$, $\sim$24\% of the planet still remains free from ice immediately around the substellar point. The addition of modest amounts of CO$_2$ and CH$_4$ appropriate for modern Earth and various Archean Earth atmospheric compositions naturally leads to warmer surface temperatures and increasingly larger fractions of open ocean. Still, even with a relatively strong Archean Earth greenhouse (0.1~bar of CO$_2$ and 10$^{-3}$~bar CH$_4$), global mean temperatures remain near $\sim$260~K, with sea ice fractions of over 60\%. H$_2$ can also increase the planetary surface temperature through greenhouse forcing from collision induced absorption with N$_2$, as proposed by \citet{wordsworth&pierrehumbert:2013}. However, our hydrogen-supporting atmosphere (Case 13) only reached a surface temperature of 267.6~K. In order for a planet with a hydrogen-supporting atmosphere to reach modern-Earth-like surface temperatures, it would require either greater than 10\% CO$_2$ and H$_2$ in a 1~bar total atmosphere, or increased surface pressures. 

A denser background N$_2$ atmosphere can significantly raise global mean temperatures via pressure broadening and lapse rate feedbacks \citep[e.g.,][]{goldblatt:2009}. With a 4~bar total atmosphere, an Archean Earth composition could raise global mean temperatures to 307~K, and the planet would be globally ocean-covered. However, with too much atmosphere, TOI-700~d could be rendered too hot for habitability. For an Archean Earth composition with a 10~bar background atmosphere, the global mean surface temperature rises to $\sim$360~K. While still technically habitable with surface liquid water and the possibility for thermophilic life forms, such a hot scenario may mark the practical upper limits on habitable atmospheric compositions for TOI-700~d. CO$_2$ dominated atmospheres, such as those expected for a habitable early Mars, can also produce broadly habitable conditions for ocean-covered planets. With 1~bar of CO$_2$, the global mean surface temperatures stabilize at a clement 284~K, and only a trace amount of sea-ice is present. Yet with a 4~bar CO$_2$ dominated atmosphere, the global mean temperature reaches $\sim$364~K, our warmest simulation. Note that none of our climates exist in a classical moist greenhouse state. Even for our warmest climate states, with T$_S$ $>$ 360~K, stratospheric water vapor remains no greater than 10$^{-4}$, as seen with the 4~bar CO$_2$ simulation, due to strong stratospheric cooling from CO$_2$ \citep{wordsworth&pierrehumbert:2013}. However, this does not rule out the possibility for water loss from exceedingly thin atmospheres (0.1 bar and less), where the near-surface water vapor concentrations can become a major constituent of the atmosphere, even at relatively cool temperatures \citep{turbet:2016}.

Assuming TOI-700~d is in synchronous rotation, the rotation period of the planet is 37.43~days, making it a slow rotator \citep{carone:2015, haqq-misra:2018}. On slowly rotating aquaplanets, strong upwelling motions tend to drive the formation of thick substellar clouds (Figure~\ref{fig:cld}), resulting in elevated planetary albedos. Note that the ``Modern Earth'' aquaplanet, the ``Archean Earth'' 1~bar aquaplanet, and the ``Early Mars'' 0.5 and 1~bar aquaplanet cases all have planetary albedos between 0.35 and 0.4, caused primarily by substellar clouds. For thicker and hotter atmospheres, however, the planetary albedo begins to drop due to increased water vapor in the atmosphere (Figure~\ref{fig:q}) and subsequent near-IR absorption of incoming stellar radiation. Our model predicts that for a synchronously rotating ocean-covered TOI-700~d, planetary albedos near $\sim$0.2 correspond to surface temperatures approaching $\sim$360~K. However, if TOI-700~d is in spin-orbit resonance, the albedo could be significantly lower and yet have a lower surface temperature ($\sim$326~K). 

We have also included a test case where we assume that TOI-700~d is in a 2:1 spin-orbit resonance. In this case the planetary rotation rate is halved. The substellar point is no longer fixed with respect to the planet surface, and significant changes occur to the planet's general circulation \citep[e.g.,][]{haqq-misra:2018}. If in a 2:1 resonance, TOI-700~d would experience significantly stronger eastward winds that advect both water vapor and clouds east of the substellar point. This phenomenon can be clearly seen in Figures~\ref{fig:q} and ~\ref{fig:cld}. Because the cloud deck is shifted east of the substellar point, the western part of the substellar hemisphere becomes relatively free from clouds, exposing more ocean. Ocean has a much lower albedo than clouds, thus the planetary albedo drops considerably to 0.131 in this case, and the surface temperature warms by more than 40~K compared to a synchronous rotator with the same atmospheric composition. While 3:2 spin-orbit resonances have been more commonly studied with 3-D climate models, here we have chosen to simulate a 2:1 resonance for TOI-700~d in order to accentuate changes to the atmospheric circulation state driven by the Coriolis force. By doubling the rotation rate of the planet compared to the synchronous case, zonal transports become enhanced \citep[e.g.,][]{kopparapu:2017,haqq-misra:2018}. Indeed, here we find a significant eastward shift of the clouds for 2:1 rotators compared to the synchronous cases. \citet{boutle:2017} similarly find that cloud and precipitation patterns are shifted slightly eastward for Proxima Centauri b, assuming a 3:2 resonance. \citet{boutle:2017} used an eccentricity of 0.3 in their resonant case, while here we have kept eccentricity at 0. The addition of a non-zero eccentricity leads to changing time and spatial patterns of stellar radiation received by the planet. A changing magnitude of incident stellar radiation could drive time variations to the cloud decks, with the clouds decks becoming thicker at perihelion and thinner at aphelion. The observational consequences of such time variations have yet to be explored. We have not explored non-zero eccentricities in this work. Note that \citet{gilbert:2020} and \citet{rodriguez:2020} have reported an eccentricity of $0.032^{+0.054}_{-0.023}$ and $0.111^{+0.14}_{-0.078}$ respectively. Thus the eccentricity of TOI-700~d may be anywhere from negligible (0.009) to significant (0.251), based on the bounds of their estimates. Still, taking the best estimated values, its eccentricity is likely much smaller than 0.3; therefore the eccentricity effects on climate demonstrated by \citet{boutle:2017} would be muted for TOI-700~d.

We have also explored several desiccated land planet scenarios. These cases as simulated are technically not habitable worlds because water has been effectively removed from the planet. However, land-covered exoplanets need not actually be fully desiccated. They could have polar, night-side, or sub-surface reservoirs of water that could drive weak hydrological cycles and support locally habitable conditions at the cold traps \citep{abe:2011, kodama:2015}. The global mean surface temperatures for desiccated planets are generally 10 to 20~K lower than for aquaplanets of the same configurations, due to the absence of any water vapor greenhouse effect. However, for dry synchronously rotating land planets, the day-to-night temperature differences can be severe (Figure~\ref{fig:ts}). 

For Figures~\ref{fig:ts},~\ref{fig:q}, and~\ref{fig:cld} we show a longitudinal slice around the equator, instead of zonal-mean results for instance, because the climates of slow and synchronous rotators are more aptly described by their day-to-night asymmetries, and because it allows us to highlight the terminator regions of the planet. The vertical dotted lines in each panel mark the terminators. Longitudinal variations in temperature and atmospheric compositions inform us about phase-dependent emission and reflection light curves. Meanwhile, the atmospheric properties at the terminators are relevant for transmission spectroscopy. For all of our simulated cases, the stratospheric temperature above $\sim$1~mbar remains near 180~K regardless of the specific atmospheric composition, surface temperature, and surface pressure. Consequently, for habitable-type atmospheres, the region of the atmosphere probed by transmission spectroscopy does not vary considerably (Figure~\ref{fig:ts}). However, note that more recent iterations of the HITRAN database have increased the shortwave absorptivity of CH$_4$, and thus the stratospheric temperatures of Archean simulations are likely underestimated \citep{Byrne&Goldblatt:2014, Byrne&Goldblatt:2015}. We find that water vapor and clouds do indeed vary substantially between cases, depending on the surface pressure, temperature, and rotation state. In Figures~\ref{fig:q} and~\ref{fig:cld}, we omit land planet cases because by definition they have no water and thus no clouds. Warmer climates tend to have more water vapor throughout the atmosphere, and are also cloudier. While high-altitude clouds located along the terminators can obscure transmission spectra \citep[e.g.,][]{fauchez:2019,suissa:2020}, day-to-night asymmetries or phase shifts in clouds could possibly be inferred from thermal emission phase curves and thus give us clues about a planet's circulation and climate state \citep{yang_j:2013, haqq-misra:2018}. The remainder of this paper is devoted to exploring how we may interpret transmission spectra and thermal emission phase curves in the context of three-dimensional cloudy atmospheres, and whether these climate states can be observed with current and future instruments. 

\subsection{Transmission Spectra}
\label{sec:spectra}
\begin{figure}
    \centering
    \resizebox{\linewidth}{!}{\includegraphics{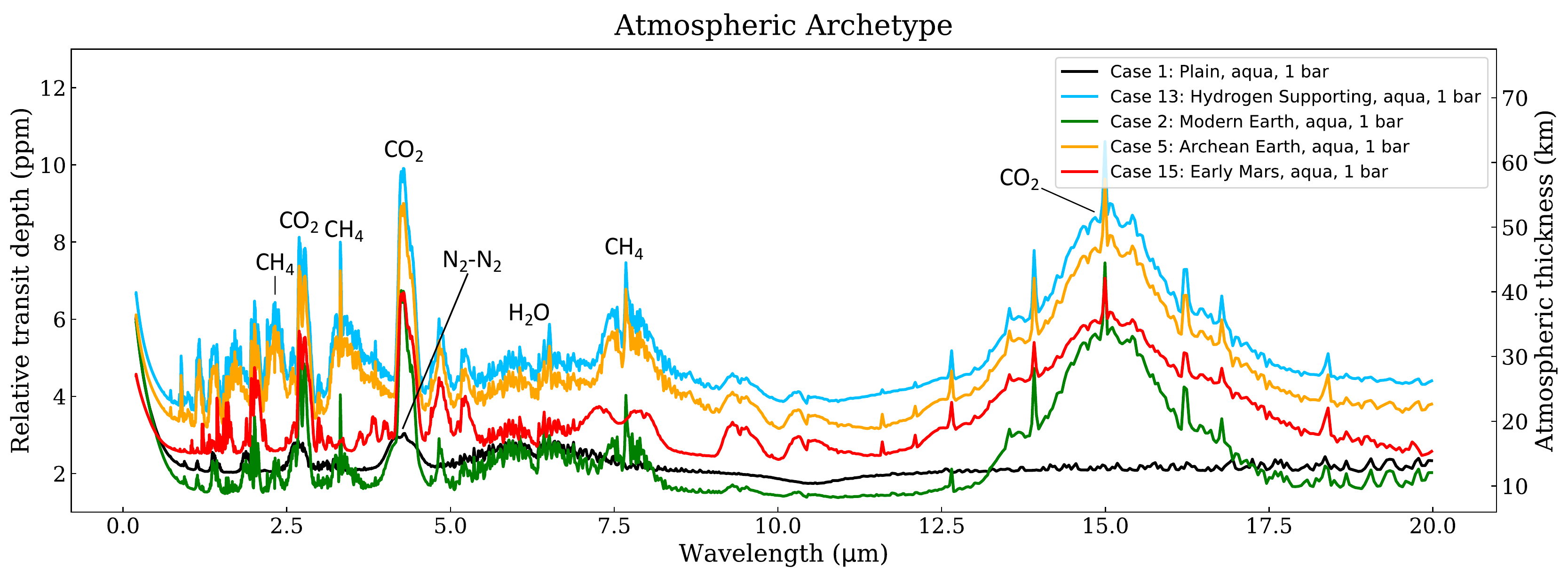}}
    \caption{Comparison of transmission spectra for different atmospheric archetypes. All simulations shown here are ocean-covered and have a surface pressure of 1~bar. Prominent features are labeled. The spectrum for the ``hydrogen-supporting'' atmosphere (10\% H$_2$) has both the highest continuum and the largest \SI{15}{\micro\metre} CO$_{2}$ feature.}
    \label{fig:archetype}
\end{figure}

\begin{figure}
    \centering
    \resizebox{\linewidth}{!}{\includegraphics{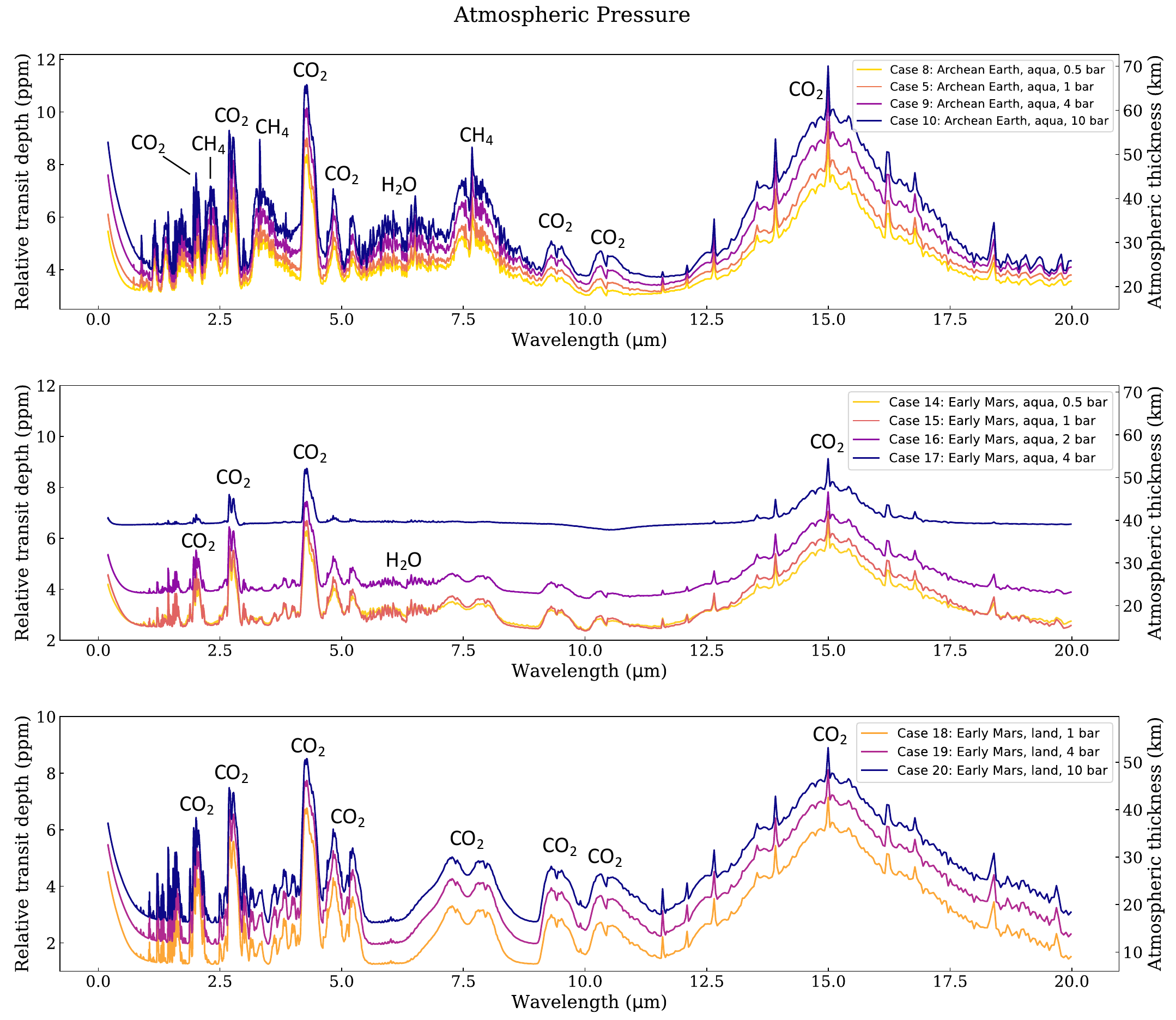}}
    \caption{Comparison of synthesized transmission spectra for different atmospheric pressures. ``Archean Earth'' aquaplanets of different surface pressures are presented in the top panel, ``Early Mars'' aquaplanets in the middle panel, and ``Early Mars'' desiccated planets in the bottom panel. Prominent features are labeled. In general, as the pressure increases, the continuum level of the spectrum increases as well.}
    \label{fig:pressure}
\end{figure}

\begin{figure}
    \centering
    \resizebox{\linewidth}{!}{\includegraphics{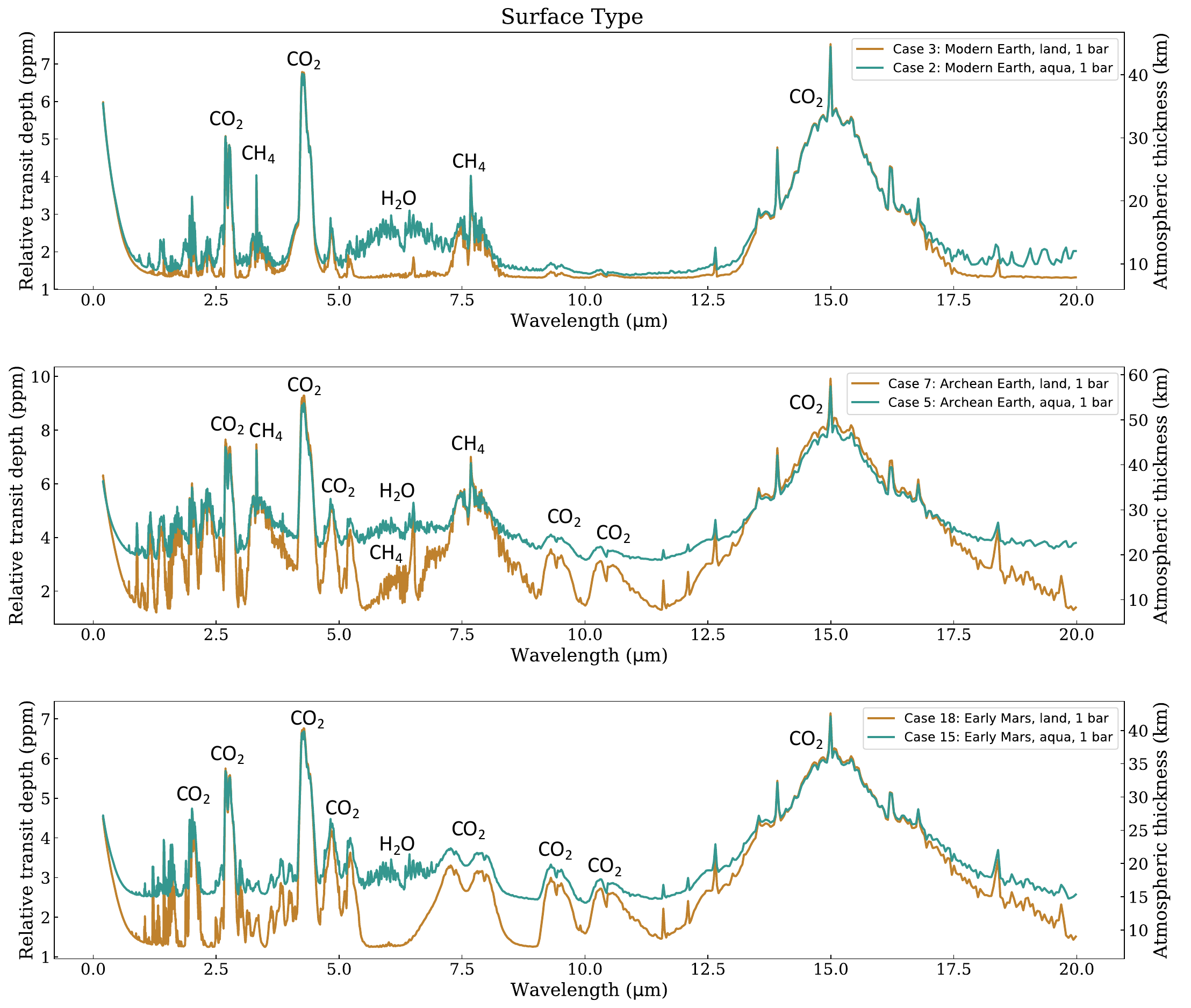}}
    \caption{Comparison of the synthesized transmission spectra of aquaplanets to their desiccated counterparts. The ocean-covered planets have higher continuum levels and reduced feature depths due to the presence of clouds. Prominent features are labeled.}
    \label{fig:surface}
\end{figure}

\begin{figure}
    \centering
    \resizebox{\linewidth}{!}{\includegraphics{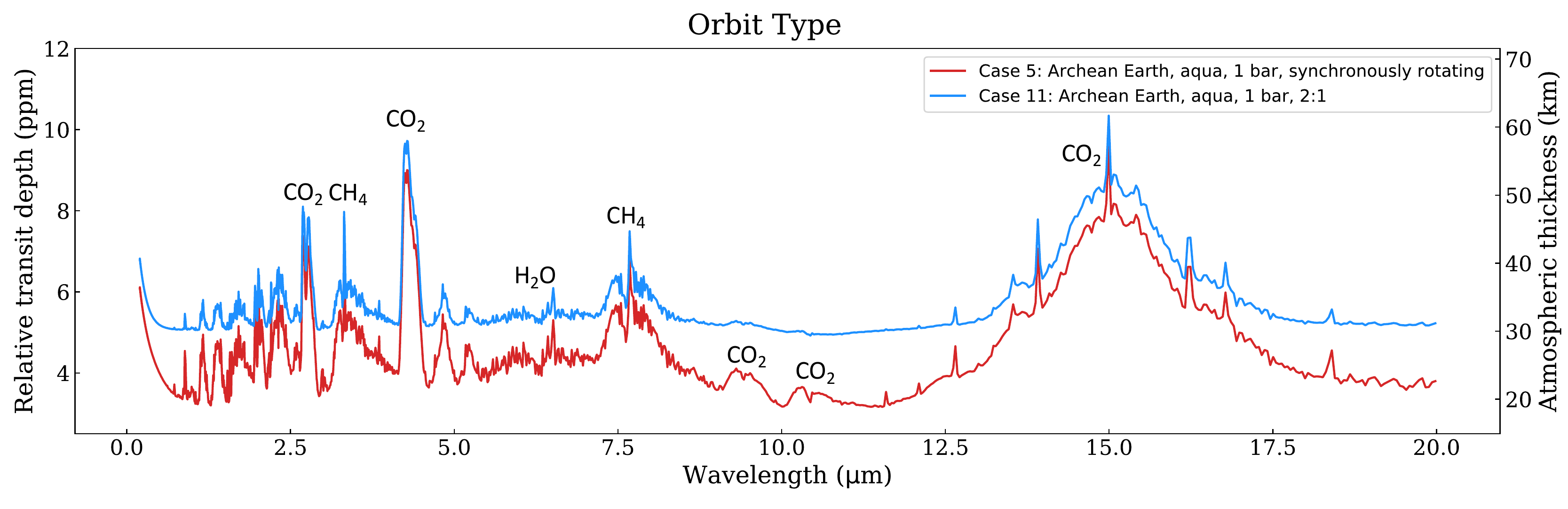}}
    \caption{Comparison of the transmission spectrum of a synchronously rotating ``Archean Earth'' aquaplanet against its 2:1 spin-orbit resonant counterpart. Although they share the same atmospheric composition, the resonant case's spectrum has a higher continuum and diminished features due to its eastern cloud formation.}
    \label{fig:orbit}
\end{figure}

We present the synthesized transmission spectra for our simulations in Figures~\ref{fig:archetype},~\ref{fig:pressure},~\ref{fig:surface}, and~\ref{fig:orbit}. Some prominent features visible in our spectra are the \SI{2.7}{\micro\metre}, \SI{4.3}{\micro\metre} and \SI{15}{\micro\metre} CO$_2$ features, the \SI{3.4}{\micro\metre} and \SI{7.6}{\micro\metre} CH$_4$ features, the \SI{1.1}{\micro\metre}, \SI{1.4}{\micro\metre} and \SI{6}{\micro\metre} water bands and the \SI{4.3}{\micro\metre} N$_2$-N$_2$ CIA. 

In Figure~\ref{fig:archetype}, we compare transmission spectra for different atmospheric compositions. We show a spectrum of each atmospheric composition archetype: ``Modern Earth'', ``Archean Earth'', and ``Early Mars'' compositions, as well as the ``plain'' aquaplanet and ``hydrogen-supporting'' atmosphere. All cases displayed in Figure~\ref{fig:archetype} are ocean-covered and have a surface pressure of 1~bar. The plain aquaplanet, void of any atmospheric constituents and spectral features apart from N$_2$ and H$_2$O, has a very shallow spectrum. Its \SI{6}{\micro\metre} H$_2$O feature has a depth less than 1~ppm, consistent with that expected for an Earth-sized planet synchronously rotating a $\sim$3500~K star \citep{suissa:2020,komacek:2020}. Without CO$_2$ in its atmosphere, the \SI{4.3}{\micro\metre} N$_{2}$-N$_{2}$ feature can be seen, indicative of a N$_2$ dominated atmosphere. For the Modern Earth and Archean Earth simulations, although both cases share the same background N$_{2}$ pressure of 1~bar, the continuum is higher in the spectrum for the Archean Earth aquaplanet. This is due to the higher concentration of CO$_{2}$, which results in more warming and hence more cloud formation. Clouds dictate where the continuum lies, as they block any incoming stellar radiation from penetrating into lower areas of the atmosphere. Note that the \SI{15}{\micro\metre} CO$_{2}$ feature is larger for the Archean Earth case, again because of a higher CO$_2$ abundance. Although the Early Mars simulation (Case 15) has the most CO$_2$ in its atmosphere, it has both a smaller \SI{15}{\micro\metre} CO$_{2}$ feature and a lower continuum than the Archean Earth case (Case 5). The feature height at \SI{15}{\micro\metre} is smaller in the Early Mars simulation because, while it does experience substantial warming (it has the highest global surface temperature out of all the simulations in Figure~\ref{fig:archetype}), it has a higher mean molecular weight, decreasing the scale height and thus the height of its spectral features. Differences in the ice cloud water abundances and ice cloud particle sizes, along with this difference in scale height, result in the Early Mars simulation having a lower continuum than the Archean Earth simulation. Keeping surface pressure and surface-type (i.e., ocean-covered) constant, it is the hydrogen-supported (10\% H$_2$) atmosphere whose spectrum has the highest continuum and the largest \SI{15}{\micro\metre} CO$_{2}$ feature. The hydrogen-supported atmosphere is identical to the Archean Earth composition, except 0.1~bar of N$_2$ is replaced with H$_2$. A planet with 10\% of its N$_2$ replaced by H$_2$ has a larger scale height; thus the spectral features for the hydrogen-supported simulation are larger than those of the Archean Earth case. In addition, the continuum is higher for the spectrum of the hydrogen-supporting planet because it is warmer; as detailed in Section~\ref{sec:climate}, H$_2$ collisions with N$_2$ produce a warming effect that is absent in the Archean Earth case. 

In Figure~\ref{fig:pressure}, we demonstrate the effect that varying background pressures have on the transit spectra. In general, the continuum level for the relative transit depth rises with increasing pressure, as can be seen distinctly in the first panel of Figure~\ref{fig:pressure}. This is because for higher surface pressures, the lower atmosphere is denser, rendering it more opaque to infrared radiation. The continuum level therefore rises above the surface to higher altitudes where the opacity is reduced. In addition, higher atmospheric pressure leads to more atmospheric refraction, also contributing to the increase in the continuum level. For all of the aquaplanet cases, clouds are present in the atmosphere. They partially or completely block the incoming stellar radiation, situating the continuum level above the cloud deck. Higher pressure atmospheres can further augment the production of high-altitude clouds by inducing a greenhouse warming effect due to pressure broadening, allowing for a warmer planet where more clouds form and push the continuum higher. In the second panel of Figure~\ref{fig:pressure}, the continuum level for the 4~bar CO$_2$ atmosphere is at $\sim$40~km, substantially higher than any of the lower pressure cases. The dramatic rise in the continuum for this particular simulation is not just due to the opaqueness of its lower atmosphere or pressure broadening. Note that in the 10~bar N$_2$ case in the top panel of Figure~\ref{fig:pressure}, clouds do not push the continuum level as high. This is because, while N$_{2}$ pressure broadening does produce a moderate greenhouse warming \citep{vonparis:2013b,wordsworth&pierrehumbert:2013,goldblatt:2009,paradise:2019}, it pales in comparison to the warming produced by a potent greenhouse gas such as CO$_{2}$, even if the N$_{2}$ atmosphere has a higher pressure than the CO$_2$ dominated planet. Since both the 10~bar N$_2$ and the 4~bar CO$_2$ simulations are ocean-covered, the simulation that experiences a greater greenhouse warming and an increased production of high-altitude water ice clouds will have the highest continuum level. For the desiccated planet cases in the bottom panel of Figure~\ref{fig:pressure}, an increase in the CO$_{2}$ pressure results in the increase of the continuum only due to the atmospheric refraction, as there are no clouds. 

Clouds not only affect the continuum level, but the depths of the individual transit spectral features themselves. For example, in the middle panel of Figure~\ref{fig:pressure}, one can notice the rapid decrease in the relative transit depth of the \SI{15}{\micro\metre} CO$_{2}$ feature as the CO$_{2}$ pressure increases. In agreement with \citet{fauchez:2019} we can see that at larger concentrations of CO$_{2}$, clouds condense at higher altitudes, thus raising the continuum to the cloud deck. As a result, the CO$_{2}$ lines shrink from the bottom leading to a decrease in their relative transit depth.

The addition of varying amounts of N$_{2}$ as a background gas introduces features that could be indicative of a dense nitrogen atmosphere \citep{komacek:2020}. For instance, the top panel in Figure~\ref{fig:pressure} shows N$_{2}$ pressure broadening features just below \SI{2.5}{\micro\metre} and around \SI{3}{\micro\metre}, both of which are absent in the CO$_2$ dominated atmospheres in the middle and bottom panels. Even as little as 0.5~bar of N$_{2}$ seems to be large enough for these features to appear in the transmission spectrum. It should be noted that nitrogen is the primary (for Earth) or the second most abundant gas (for Venus and Mars) in the terrestrial planetary atmospheres in our Solar System. If this is true for exoplanetary atmospheres as well, N$_{2}$ pressure broadening features could be a marker for such atmospheres. In addition, N$_{2}$-N$_{2}$ collision-induced absorption (CIA) at $\sim$\SI{4.3}{\micro\metre}, in the wing of the CO$_2$ absorption feature, can be another indication of a N$_{2}$-rich atmosphere as shown by \citet{Schwieterman:2015}. In the absence of CO$_2$, this signal could be substantially larger, especially for planets with significant H$_2$ mixing ratios.

In Figure~\ref{fig:surface}, we compare the spectral feature differences between ocean-covered worlds and desiccated planets. In each panel it is apparent that the aquaplanet has a higher continuum level as well as reduced feature depths compared to its land-covered counterpart, due to the presence of high-altitude water ice clouds. Likewise, the spectra for the ocean-covered planets include water vapor features, such as the \SI{6}{\micro\metre} feature, which are absent in the dry cases. This absence reveals that the dry 1~bar ``Archean Earth'' case has enough methane in its atmosphere to have a \SI{6}{\micro\metre} CH$_4$ feature, which could be useful in identifying an atmosphere as dry, haze-free and methane-rich. To distinguish between desiccated planets that have ``Modern Earth''-like and ``Archean Earth''-like compositions, identifying CH$_{4}$ features can be a key observational constraint. The strength of the CH$_{4}$ features increases with the increase of the CH$_{4}$ abundance in the atmosphere, as can be seen by comparing the heights of the $\sim$\SI{3.7}{\micro\metre} or the $\sim$\SI{7.5}{\micro\metre} features between the ``Archean Earth'' and ``Modern Earth'' cases. Note that our ``Early Mars'' atmospheres only have CO$_2$ and H$_2$O; CH$_4$ thus is not present in their associated transmission spectra. Figure~\ref{fig:surface} also indicates that, while the \SI{15}{\micro\metre} CO$_{2}$ feature, irrespective of its abundance, may betray the presence of an atmosphere, its detection alone cannot specify an atmospheric archetype or surface condition, because it is prominent in all of the simulated cases (except for the ``plain'' aquaplanet and the 4~bar CO$_2$ dominated aquaplanet in the middle panel of Figure~\ref{fig:pressure}).

We highlight the effect that different orbit phenomena have on transmission spectra in Figure~\ref{fig:orbit}. The ``Archean Earth'' aquaplanet in a 2:1 rotation:orbit resonance has a higher continuum and smaller spectral feature depths than its synchronously rotating counterpart does, although the atmospheric compositions are the same. This is because, as seen for Case 11 in Figure~\ref{fig:cld}, the resonant simulation experiences an eastward shift of large clouds. These clouds on the eastern side of the planet breach and heavily obscure the terminator region, the area of the planet sensitive to transmission spectroscopy. When comparing the ``Archean Earth'' 1~bar land planet against its resonant desiccated analogue, the transmission spectra (not shown here) are identical due to the lack of the clouds that produce this effect. 

A property shared by all of our generated transmission spectra, including those displayed in Figure~\ref{fig:pressure},~\ref{fig:surface},~\ref{fig:orbit}, is that the spectral feature depths are low, with the relative transit depth for the large \SI{15}{\micro\metre} CO$_2$ feature usually hovering between 4 and 6~ppm (with the exception of the 4~bar CO$_2$ dominated case, for which the \SI{15}{\micro\metre} CO$_2$ feature has a depth of $\sim$1~ppm, and the ``plain'' aquaplanet, which has no CO$_2$). The implications of these low signals are discussed in Section~\ref{sec:discussion}. 

\subsection{Combined Phase Curves}
\label{sec:phasecurves}

\begin{figure}
    \centering
    \begin{subfigure}{}
        \centering
        \includegraphics[width=\textwidth]{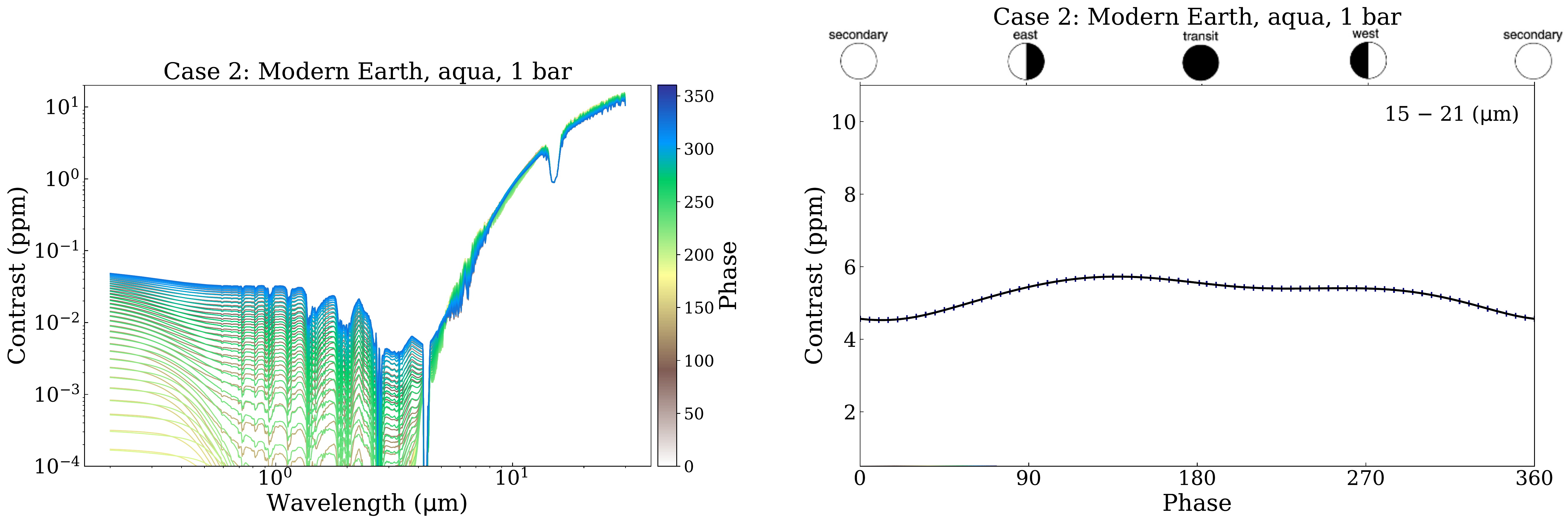}
        \label{fig:modernearth_aqua}
    \end{subfigure}
    \begin{subfigure}{}
        \centering
        \includegraphics[width=\textwidth]{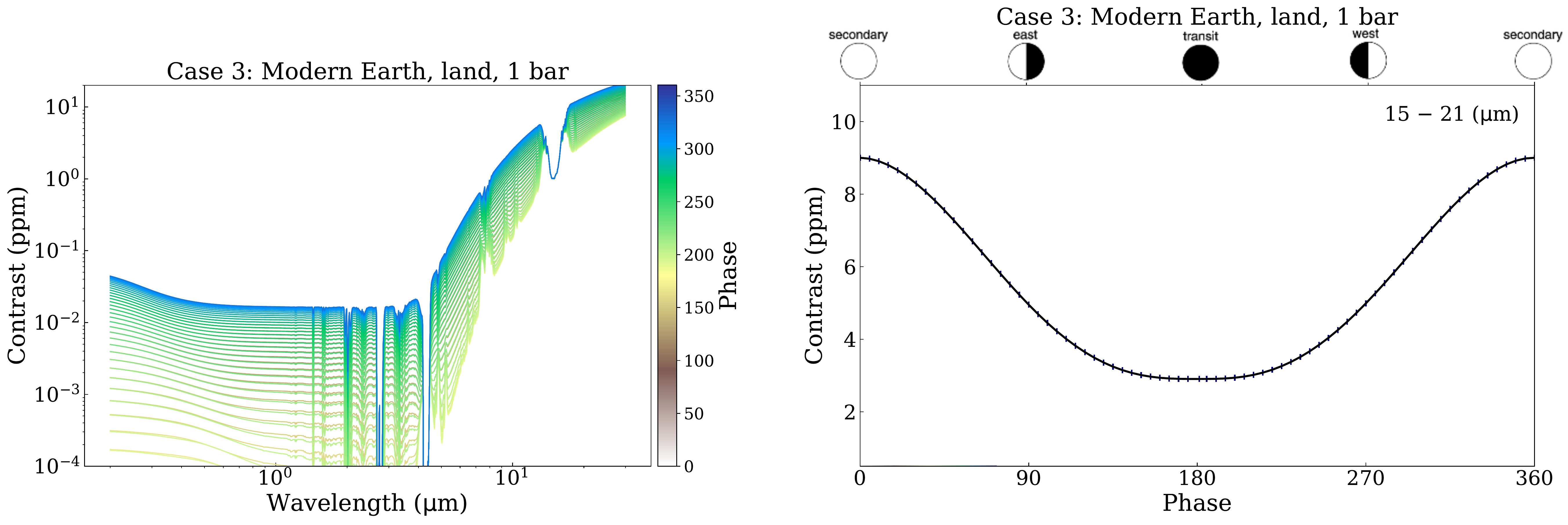}
        \label{fig:modernearth_land}
    \end{subfigure}
    \caption{Combined-light phase-dependent spectra (left panel) and integrated broadband phase curves (right panel) for the Modern Earth aquaplanet (top) and the Modern Earth desiccated planet (bottom). Transit occurs at $180^{\circ}$. In the left panel, the contribution of reflected light dominates at wavelengths $<\sim$\SI{4.3}{\micro\metre}, while thermally emitted light dominates at wavelengths $>\sim$\SI{4.3}{\micro\metre}. In the right panel, the integrated fluxes for the phase curves are $15 - $\SI{21}{\micro\metre}.}
    \label{fig:modern_phase}
\end{figure}

\begin{figure}
    \centering
    \begin{subfigure}{}
        \centering
        \includegraphics[width=\textwidth]{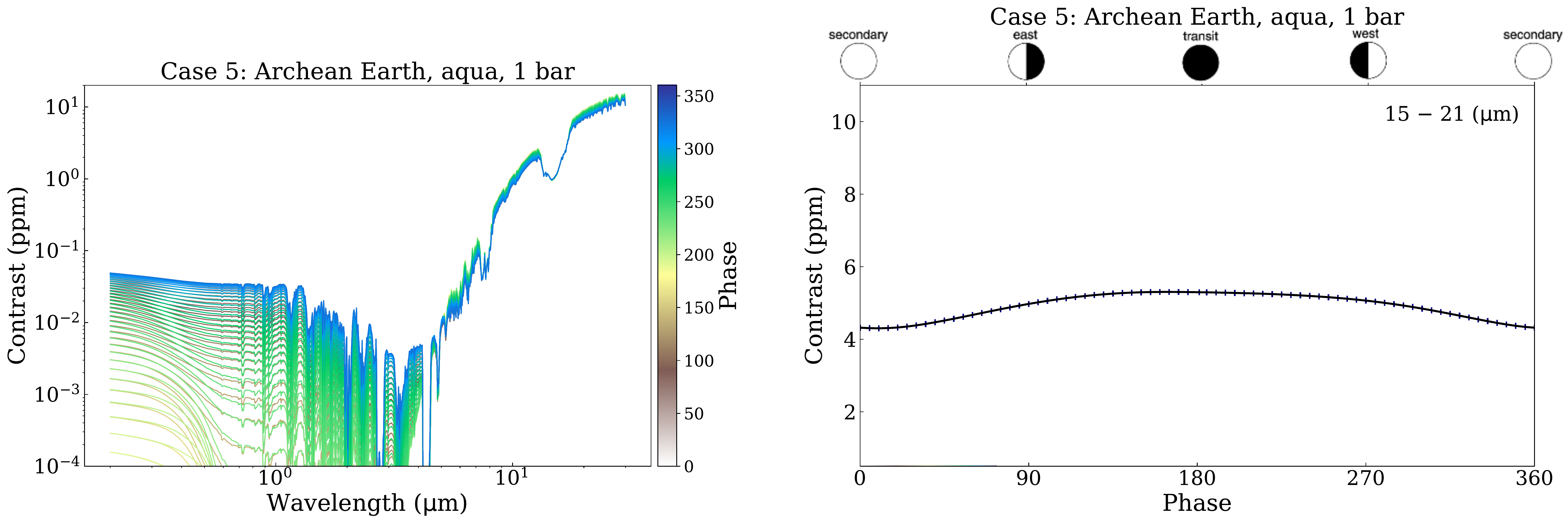}
        \label{fig:archean2_1bar_aqua}
    \end{subfigure}
    \begin{subfigure}{}
        \centering
        \includegraphics[width=\textwidth]{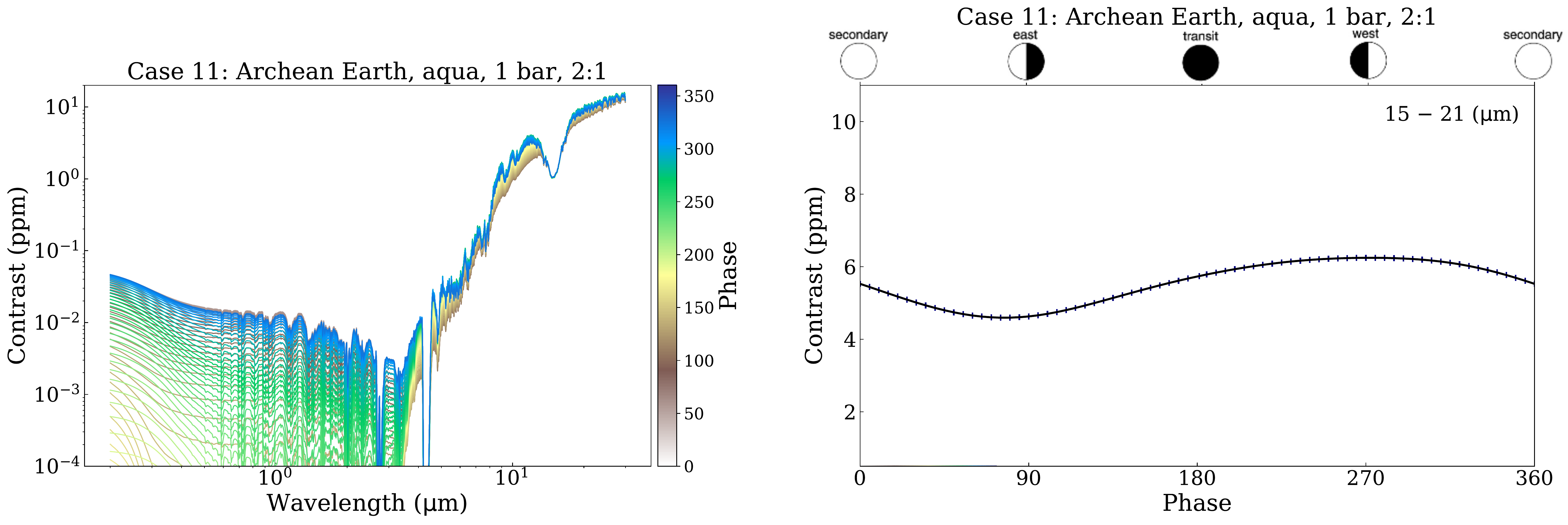}
        \label{fig:archean2_1bar_so21}
    \end{subfigure}
    \caption{Combined-light phase-dependent spectra (left panel) and integrated broadband phase curves (right panel) for a synchronously rotating 1~bar Archean Earth aquaplanet (top) and its 2:1 spin-orbit resonant aquaplanet counterpart (bottom). Transit occurs at $180^{\circ}$. In the left panel, the contribution of reflected light dominates at wavelengths $<\sim$\SI{4.3}{\micro\metre}, while thermally emitted light dominates at wavelengths $>\sim$\SI{4.3}{\micro\metre}. In the right panel, the integrated fluxes for the phase curves are $15 - $\SI{21}{\micro\metre}.}
    \label{fig:orbit_phase}
\end{figure}

\begin{figure}
    \raggedleft
    \begin{subfigure}{}
        \raggedleft
        \includegraphics[width=0.9\textwidth]{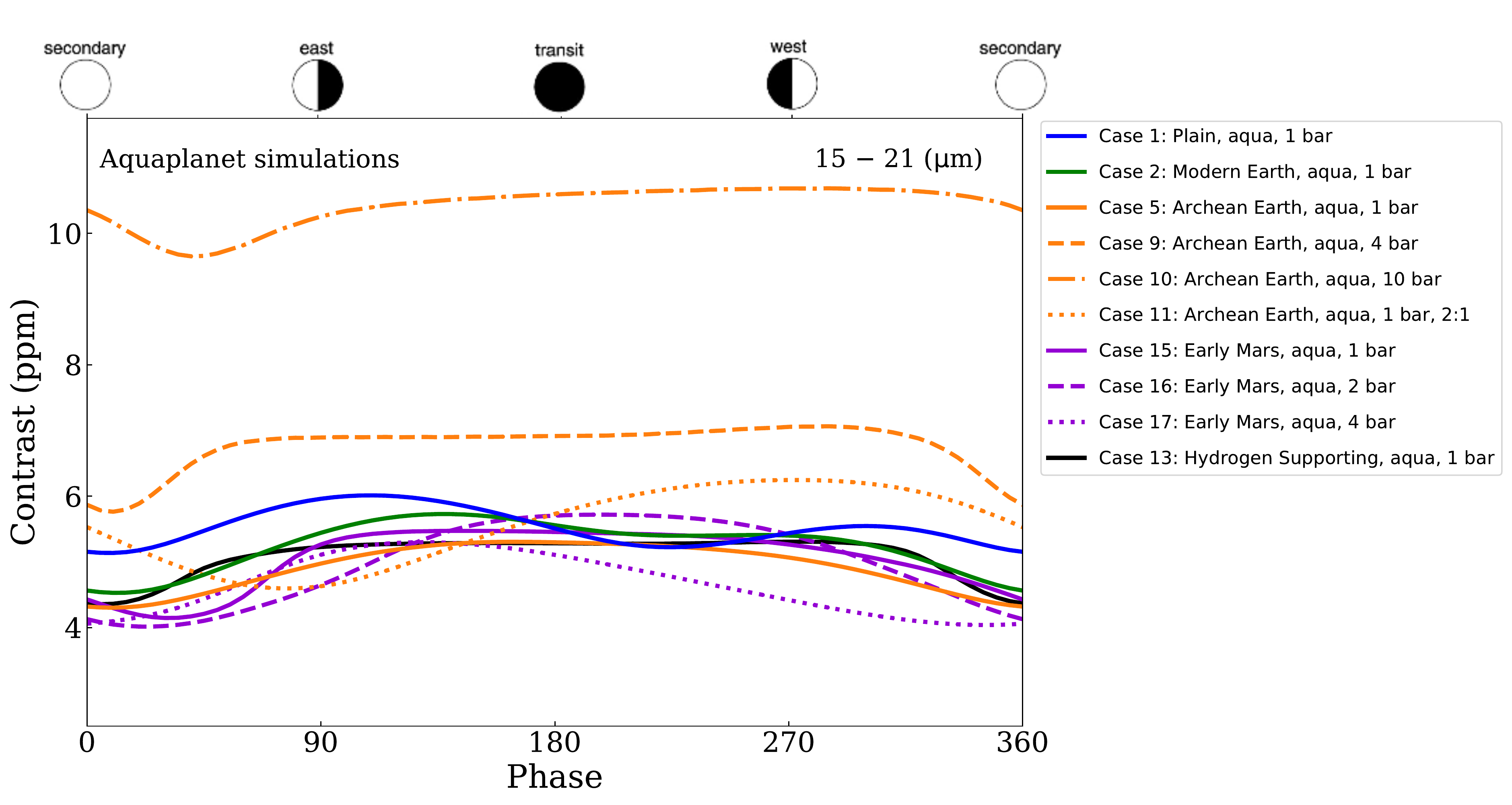}
        \label{fig:aqua_all}
    \end{subfigure}
    \begin{subfigure}{}
        \raggedleft
        \includegraphics[width=0.9\textwidth]{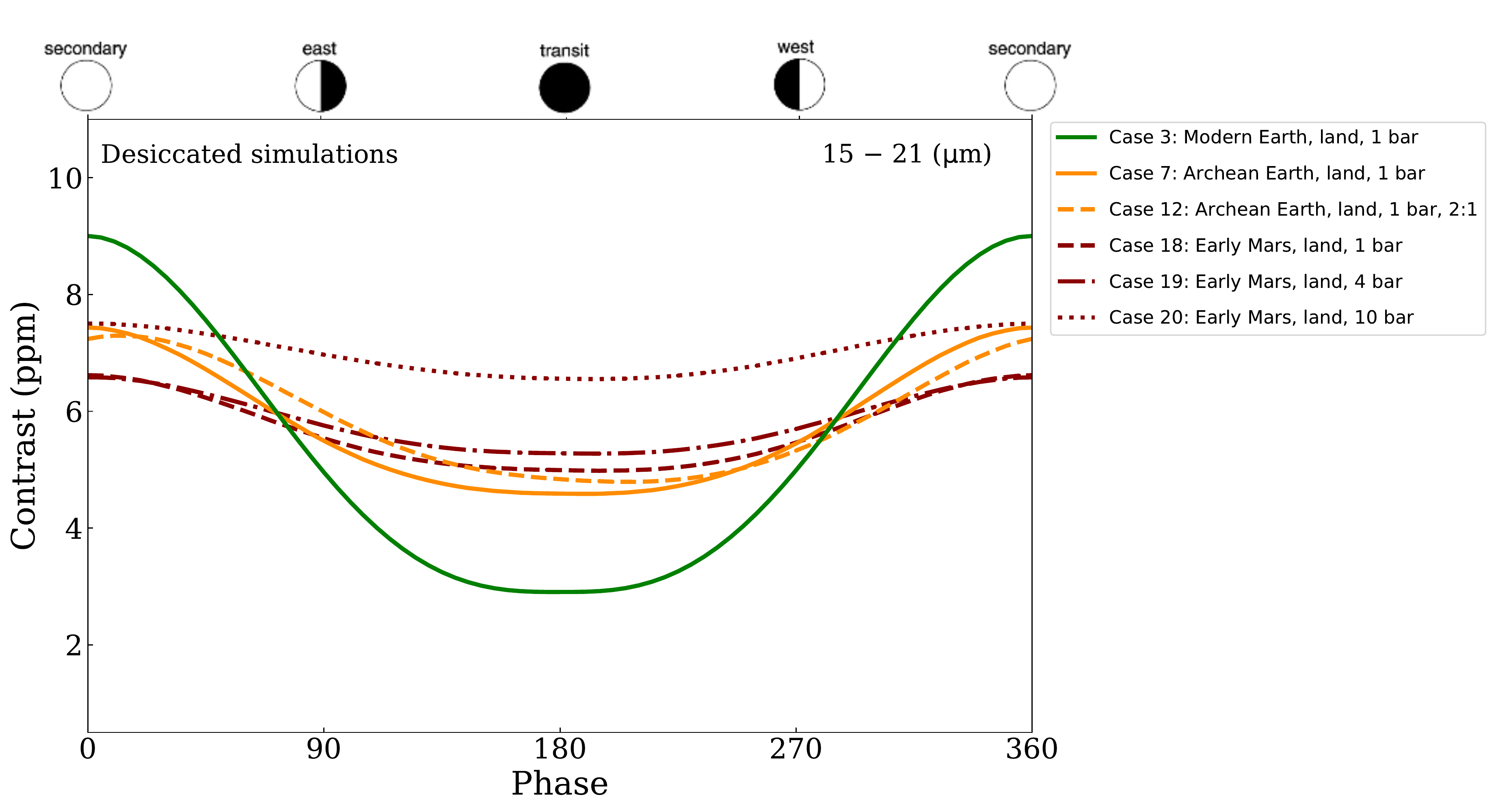}
        \label{fig:land_all}
    \end{subfigure}
    \caption{Phase curves for the majority of our simulations in this work. Aquaplanets are shown in the top panel; desiccated planets are shown in the bottom panel. Transit occurs at $180^{\circ}$. In the left panel, the contribution of reflected light dominates at wavelengths $<\sim$\SI{4.3}{\micro\metre}, while thermally emitted light dominates at wavelengths $>\sim$\SI{4.3}{\micro\metre}. In the right panel, the integrated fluxes for the phase curves are $15 - $\SI{21}{\micro\metre}.}
    \label{fig:all_phase}
\end{figure}

In Figures~\ref{fig:modern_phase},~\ref{fig:orbit_phase}, and~\ref{fig:all_phase}, we present combined-light (thermal emission and reflected light) phase-dependent spectra (left panel) and integrated broadband phase curves (right panel) for different atmospheric compositions of TOI-700~d. Note that for the broadband phase curves, a phase of $180^{\circ}$ is the transit and $0^{\circ}$ is the secondary eclipse. The integrated fluxes are from the $15 - $\SI{21}{\micro\metre} bandpass, where the thermal emission from the planet is dominant. The morphology of a broadband phase curve is driven by longitudinal variations, which result from both the atmospheric composition and dynamics \citep{selsis:2011}. For the aquaplanet cases, the shape of the phase curve traces the extent of the cloud cover on the planet. This aspect highlights the advantage of using a 3-D climate model, where the global cloud distribution can be self-consistently simulated to assess its effect on observables. 

Figure~\ref{fig:modern_phase} shows the distinct variation in both the spectrally resolved and broadband phase curves between ocean-covered (top) and desiccated (bottom) planet configurations, assuming 1~bar ``Modern Earth'' atmospheric composition for TOI-700~d. Several interesting features can be noted comparing the two cases (Cases 2 and 3 in Table~\ref{tab:climate_properties}). The land planet simulation has an unmistakable minimum at the transit phase ($180^{\circ}$). Meanwhile, the aquaplanet model has a mostly flat phase curve, with minima at the secondary eclipse phases ($0^{\circ}$ and $360^{\circ}$), and perhaps a very slight dip close to transit. The morphology of the phase curve for the desiccated planet effectively illustrates that there is a larger day-to-night contrast in the desiccated planet case compared to its flat aquaplanet counterpart. This difference in phase curve morphology between aquaplanets and desiccated planets is also broadly consistent with the comparisons made in \cite{turbet:2016} (see their Figures B23 and B20). Without water vapor to efficiently transport heat and without clouds to influence thermal emission \citep[e.g.,][]{yang_j:2013}, the locked dayside of the land planet receives and reflects most of the available flux from the host star, and also has the highest temperatures and thermal flux emitted to space. The land planet's phase curve distinguishes itself from that of a bare rock as it has a lower amplitude than what is empirically expected for an atmosphere-less planet \citep{selsis:2011,maurin:2012,koll:2015}, due to the existing heat distribution of the atmosphere. For the aquaplanet case in Figure~\ref{fig:modern_phase}, the minima at the secondary eclipse are due to the substellar clouds, which inhibit the escape of long-wave thermal radiation \citep[consistent with Figure 3 in][]{yang_j:2013}. As there are no clouds in the land planet case, no such reduction can be seen in the broadband phase curve near the secondary eclipse. 

The phase-dependent combined-light spectra in Figure~\ref{fig:modern_phase} (left panel) also show distinct variations, particularly in the near-IR ($\sim$0.8$ - $\SI{3}{\micro\metre}) for the aquaplanet case, where water vapor absorption can be identified. In both of these cases, the strong \SI{15}{\micro\metre} CO$_{2}$ absorption feature stands out, betraying the presence of an atmosphere. While transit observations can potentially detect the presence of CO$_{2}$ in the atmosphere in the near-IR part of the spectrum, thermal emission spectra can also be used to verify the existence of a CO$_{2}$ dominated atmosphere.

In Figure~\ref{fig:orbit_phase} we highlight the effect that rotation has on phase curve morphologies. The top panel is a synchronously rotating 1~bar aquaplanet with an ``Archean Earth'' composition (Case 5 in Table~\ref{tab:climate_properties}). The bottom panel is the aquaplanet simulation with the same atmospheric properties, except in a 2:1 rotation:orbital resonance (Case 11). The synchronously rotating case has a similar phase curve morphology as that of the ``Modern Earth" aquaplanet displayed in Figure~\ref{fig:modern_phase}. However, the ``Archean Earth'' is slightly flatter, corresponding to less of a day-to-night temperature variation, due to the increased amount of CO$_2$ warming in the atmosphere. Meanwhile, the 2:1 resonant case in Figure~\ref{fig:orbit_phase} dips when the east side of planet comes into view. As can be seen in the cloud profile for this case in Figure~\ref{fig:cld}, there is heavy cloud cover asymmetrically skewed towards the east side of the planet. The eastern clouds block the outgoing infrared radiation, resulting in a low contrast of thermal emission at $90^{\circ}$. As the planet approaches transit and beyond, the thermal emission increases due to the westward dwindling of clouds. There is no minimum of thermal emission at transit, because the planet is not tidally locked, and thus the longitudinal distribution of temperature is more or less uniform. The only factor that contributes to the variations in the resonant case's phase curve is the distribution of clouds. 

We overlay synthesized phase curves for the majority of our simulations in Figure~\ref{fig:all_phase}, which is divided into aquaplanets (top panel) and desiccated planets (bottom panel). The 10~bar ``Archean Earth'' simulation (Case 10) has the highest contrast ($\sim$10~ppm) due to the combination of its high temperatures and relatively mild cloud coverage. Although the 4~bar ``Early Mars'' planet (Case 17) is our hottest simulation, it has a substantial amount of clouds (see Figure~\ref{fig:cld}), and thus has a lower-contrast phase curve than the 10~bar ``Archean Earth'' planet. The bottom panel of Figure~\ref{fig:all_phase} demonstrates that the ``Modern Earth'' desiccated case not only has the largest day-to-night temperature differences, but also has the second largest contrast of our simulated phase curves ($\sim$9~ppm). All of our simulations yield maximum variation below 10~ppm, with the average maximum contrast being $\sim$6~ppm. The observability prospects for these low signals are discussed in Section~\ref{sec:discussion}.

\section{Discussion}
\label{sec:discussion}
In this paper, we have explored possible habitable climates for the first Habitable Zone Earth-sized planet discovered by the \textit{TESS} mission: TOI-700~d \citep{gilbert:2020,rodriguez:2020}, which serves as both a potential target itself for future characterization and also as a benchmark planet for understanding the potential climate states of Earth-sized planets in the Habitable Zones of early M-dwarfs. We have presented synthesized transmission spectra, phase-dependent combined-light spectra, and integrated broadband phase curves for the planet, based on ExoCAM GCM simulations of a diversity of climate conditions. Receiving $\sim$0.86~S$_0$ from a $\sim$3500~K star makes TOI-700~d a strong candidate for a habitable world. As outlined in Section~\ref{sec:climate}, for the cases where we assume the planet is an ocean world and in synchronous rotation, we find that the planet still maintains habitable conditions for a diverse set of surface pressures and atmospheric conditions, with our coldest simulation (the ``plain'' aquaplanet) still maintaining $\sim$24\% of open ocean. We use the 10~bar ``Archean Earth'' and 4~bar ``Early Mars'' cases to mark the upper limits of habitability for the simulations we explore here for TOI-700~d. For desiccated planets, the global mean surface temperatures are generally 10 to 20~K lower than for aquaplanets of the same configurations because of the absence of any water vapor. For the non-synchronous, 2:1 spin-orbit resonance aquaplanet, we find that the surface temperature warms by more than 40~K compared to a synchronous case with the same atmospheric composition. We also have presented several ways to disentangle the different climate states from the observables we synthesized. It may be possible, for example, to distinguish a dry, haze-free, methane-rich atmosphere, as well as discern between N$_2$ and CO$_2$ dominated atmospheres. The distinctions between phase curve morphologies for desiccated and ocean-covered worlds, as well as a synchronously rotating aquaplanet and a 2:1 resonant aquaplanet are also quite clear. Our observables show that the presence of a CO$_{2}$-containing atmosphere can be easily discernible via CO$_{2}$ spectral features, provided our instruments capabilities allow it (see Section~\ref{sec:prospects}). 

\subsection{Comparison with Previous GCM Studies of Terrestrial Planets in the Habitable Zone}
\label{sec:otherplanets}
Although our work is specific to the parameters of TOI-700~d, our study builds upon previous GCM studies that have been used to model other terrestrial-sized exoplanets in the habitable zones of M dwarf stars. For instance, \citet{wordsworth:2011} studied the habitability of the now non-existent exoplanet Gliese 581~d \citep{robertson:2014}. Gliese 581~d \citep{udry:2007} was claimed to lie just outside the outer edge of the conservative HZ \citep{selsis:2007,chylek&perez:2007,von-bloh:2007}. \citet{wordsworth:2011} performed GCM simulations with varying CO$_2$ abundances, background pressures, and rotation resonances for both desiccated and aquaplanets. They concluded that for atmospheres with more than 10 bars of CO$_2$, Gl 581~d could maintain surface liquid water. This requirement of a thick atmosphere for habitability is not necessary for TOI-700~d as it lies comfortably within the conservative HZ, receiving $\sim$87\%~S$_0$, equal to the solar flux Earth received $\sim$1.7 billion years ago \citep{gough:1981} and with a redder stellar spectral energy distribution which is more readily absorbed by atmospheric gases and surface materials compared to the Sun’s energy.

TOI-700~d is perhaps more comparable with Proxima Centauri~b \citep{anglada-escude:2016} which receives $\sim$65\%~S$_0$, or TRAPPIST-1~e \citep{gillon:2016,gillon:2017} which receives $\sim$66\%~S$_0$. Both of these worlds have been studied at length with multiple GCMs. The nearby, non-transiting Proxima Centauri~b has been simulated by \cite{turbet:2016,boutle:2017,delgenio:2019} using GCMs and assuming a planetary radius of 1.1~\rearth, comparable to the radius used here for TOI-700~d. Compared to Proxima Centauri (3050~K), TOI-700 is a hotter host star (3480~K), and planet d has a longer period (37.43 Earth days) and thus the rotation rate is slower than for Proxima Centauri~b (11.2 Earth days) assuming synchronous conditions \citep{anglada-escude:2016}. As a result, TOI-700~d’s substellar cloud would be more pronounced than Proxima Centauri~b’s due to TOI-700~d’s weaker Coriolis force, which results in a thermally direct circulation with strong upwelling and cloud formation across the substellar hemisphere \citep{haqq-misra:2018}. The combination of a hotter host star and thicker clouds would result in TOI-700~d having a higher albedo than Proxima Centauri~b. The TRAPPIST-1 planets have also been studied with GCMs \citep{wolf:2017,turbet:2018,fauchez:2019}; however they orbit a much cooler star \citep[2516~K,][]{vangrootel:2018} with much shorter orbital periods and thus faster rotation rates assuming tidal synchronization. TRAPPIST-1~d and f, residing in the optimistic HZ, would be much hotter and colder respectively, compared to TOI-700~d as a simple function of their received stellar flux \citep[114\%~S$_0$ and 38\%~S$_0$,][]{gillon:2017}. While TRAPPIST-1~e makes for a closer comparison, its climate would be vastly different from TOI-700~d, due to its significantly faster rotation rate (6.1 Earth days) and extremely red incident stellar flux. The relatively rapid rotation of TRAPPIST-1~e results in strong upper-level winds which advect substellar clouds eastward of the substellar point, all but negating the substellar albedo effect seen in simulations of slower rotators such as TOI-700~d. Additionally, enhanced zonal atmospheric heat redistribution would make the night side of TRAPPIST-1~e warmer than TOI-700~d, if both had cold or temperate climate states.

Due to its longer-period orbit, TOI-700~d in some ways may be better suited for habitability than both TRAPPIST-1~e and Proxima Centauri~b. For instance, when comparing 1~bar Modern Earth aquaplanet cases simulated by GCMs without ocean heat transport for all three planets, we find that TOI-700~d has the greatest fraction of open ocean (28.8\% for Case 2 in this work, compared to 20\% for both TRAPPIST-1e in \citealt{fauchez:2019} and Proxima Centauri~B in \citealt{delgenio:2019}). However, we note that \citet{delgenio:2019} demonstrates that the fraction of open ocean for Proxima Centauri~b with a 1~bar Modern Earth atmosphere with ocean heat transport raises to 40\%. The impact that ocean heat transport has on TOI-700 d’s potential for habitability has not yet been explored.

In addition, it is important to note that cooler M-dwarfs such as TRAPPIST-1 are highly active; therefore it remains to be seen how the comparatively low activity of TOI-700 affects the atmospheric retention, and ultimately the climates, of the planets in the system. We continue to discuss the issue of atmospheric retention regarding TOI-700~d in Section~\ref{sec:atmloss}.

\subsection{Atmospheric Retention}
\label{sec:atmloss}
Although we simulate many different climatological scenarios, the extent to which TOI-700~d has suffered from atmospheric loss is an open question. The host star is a quiet early M dwarf with an age estimate of $>$1.5~Gyr \citep{gilbert:2020}. Recent studies have pointed out that lower mass stars undergo an extended phase of high-luminosity pre-main sequence evolution until they start to fuse hydrogen into helium in their cores \citep{chabrier:1997,baraffe:1998,baraffe:2015}. For a G-dwarf star like our Sun, this pre-main sequence phase lasts only $\sim$50~Myr \citep{baraffe:2015}, but the stage can last anywhere from $\sim$100~Myr to $\sim$1~Gyr for early and late M-dwarfs, respectively. TOI-700 is an early M dwarf with a mass estimate of $0.416$~\msun, and so it should have a relatively short high-luminosity pre-main sequence evolution, perhaps only $\sim$200~Myr \citep{luger&barnes:2015}. For TOI-700, the variation in the bolometric luminosity itself is less than an order of magnitude while entering into the main sequence, while a late M-star spectral type (like TRAPPIST-1) experiences several orders of magnitude variation in the latter case. This could potentially be advantageous in terms of retaining an atmosphere for the terrestrial-sized planet in the HZ, TOI-700~d. 

\begin{figure}
    \centering
        \centering
        \includegraphics[width=0.7\textwidth]{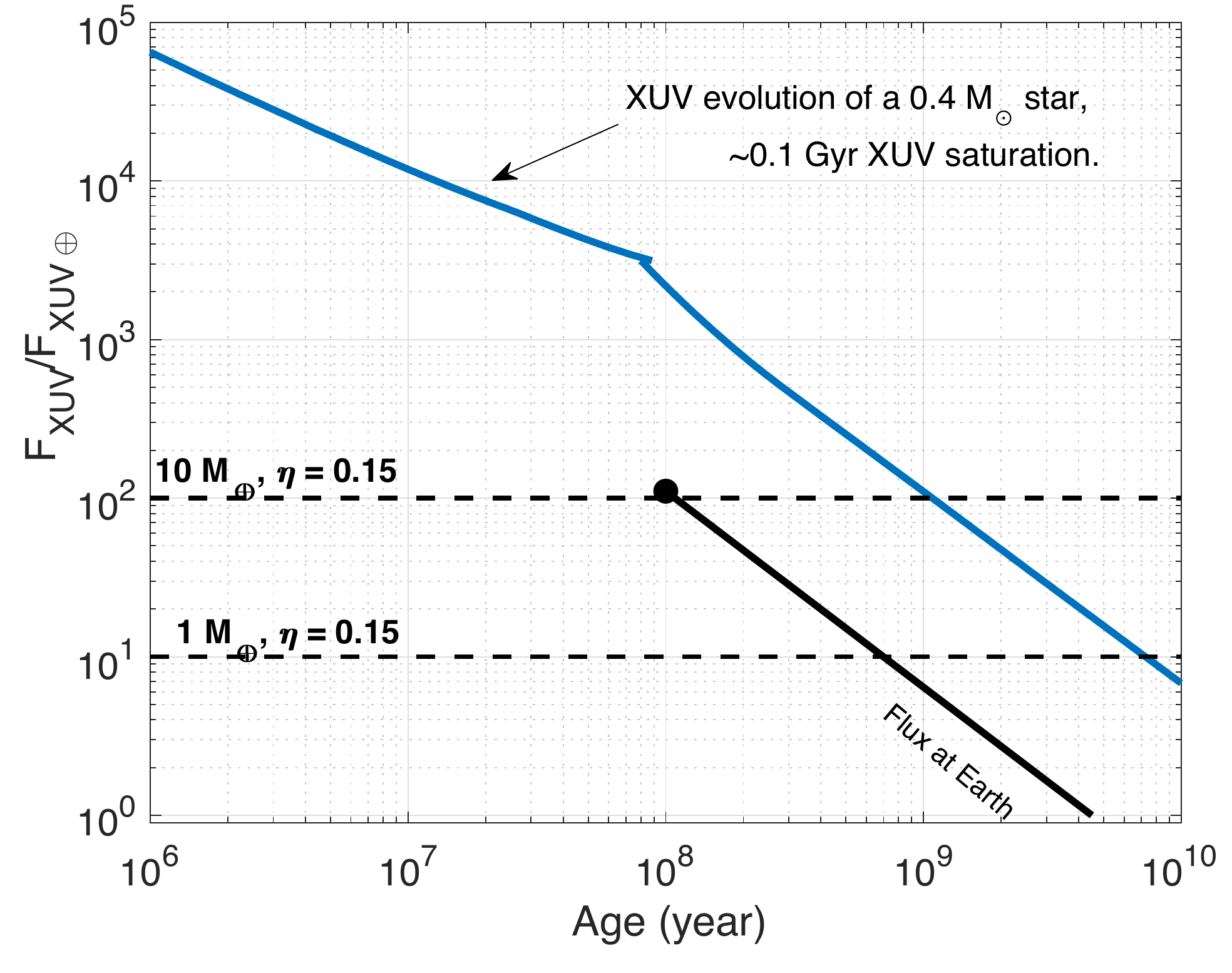}
    \caption{Incident XUV ($1 - 1000$~\si{\angstrom}) flux evolution normalized to the present Earth value ($4.64$ ergs$\cdot$s$^{-1}\cdot$cm$^{-2}$) on TOI-700~d from a star with a mass equivalent to that of its host star (blue curve). An XUV saturation time of $\sim$0.1~Gyr is assumed. The flux evolution on Earth from a Sun-like star is also shown (black solid), and the black dot represents the earliest time \citet{ribas:2005} have data for solar-type stars. The horizontal dashed lines show the threshold XUV limit for the ``blowing off'' of an atmosphere on a 1 M$_{\oplus}$ and 10 M$_{\oplus}$ planet above which the escape of the atmosphere is dominated by the hydrodynamic wind, assuming heating efficiencies of 0.15. Depending upon the age of the system, TOI-700~d may be below the threshold for the blow-off of its atmosphere.} 
    \label{xuv}
\end{figure}

However, while TOI-700 is a quiet M dwarf, this stellar type is more active than solar-type stars \citep{scalo:2007}, with X-ray and UV radiation (XUV, $1 - 1000$~\si{\angstrom}) particularly excessive during the young, pre-main sequence period of M-dwarf stars. XUV radiation can lead to atmospheric escape, stripping the atmosphere of a planet in the HZ \citep{watson:1981,lammer:2003,lammer:2009,lammer:2013,yelle:2004,erkaev:2007,erkaev:2013,tian:2009,owen&jackson:2012}. In Figure~\ref{xuv}, we compare the incident XUV radiation over time on TOI-700~d using a star with an equivalent mass of TOI-700 (solid blue line) based on the evolutionary path for XUV radiation from \citet{ribas:2005}. We assume a XUV ``saturation'' timescale (i.e, the XUV luminosity is saturated, and remains constant for 0.1~Gyr, after which it decreases) of $\sim$0.1~Gyr. For comparison, the XUV flux on Earth from the Sun is also shown (solid black line). The solid filled circle is the earliest time for which \citet{ribas:2005} have integrated XUV flux estimates for solar type stars in the wavelength interval $1 - 1180$~\si{\angstrom} (see their Table 4, entry for EK Dra). We also show the threshold XUV limit for the ``blowing off'' of an atmosphere on a 1 M$_{\oplus}$ and 10 M$_{\oplus}$ planet (dashed horizontal lines), above which the escape of the atmosphere is dominated by the hydrodynamic wind. We assume for heating efficiencies of 0.15. The heating efficiency $\eta$ is defined as the percentage of incoming XUV energy that is transferred locally into heating of the gas \citep{erkaev:2013}\footnote{The XUV heating efficiency factor in reality might vary with CO$_{2}$ abundance, as described in \citet{johnstone:2018}. However, to account for it, a complete thermal structure model of the atmosphere is needed which is beyond the scope of this paper. As such, our estimate is a first order estimate and should be considered as a preliminary step for a more comprehensive model exploration}. Given the above assumptions, Figure~\ref{xuv} indicates that Earth-mass planets around TOI-700-type stars may be in the blow-off regime for several billion years, resulting in the loss of their primary atmospheres. TOI-700~d is a planet with a mass that is likely less than $5$~M$_{\oplus}$. While a mass larger than that of the Earth could moderately reduce the atmospheric blow-off time assuming the same heating efficiency, if the planet formed with its current size, it is still likely that the planet may have lost its primordial atmosphere. 

Although there is a possibility that a primordial atmosphere may have been stripped off due to the above mentioned high luminosity pre-main sequence evolution of the host star, secondary atmospheres may build up past this phase through volcanic outgassing \citep{wordsworth&pierrehumbert:2013,driscoll:2014}. Tectonic activity (either through tidal forces, or similar to the Earth) may facilitate an efficient water cycling between the interior and the surface, potentially sustaining oceans on the surface of the planet \citep{sandu:2011,cowan:2014,schaefer:2015,komacek&abbot:2016}. The survival of oceans on the surface of an M-dwarf planet depends upon the stellar UV activity, and the corresponding escape of hydrogen into space. Although we did not include the effect of photochemistry for our climate simulations in this work, other studies have used either a self-consistent GCM-photochemistry model to directly simulate the resultant H$_2$ mixing ratios due to stellar activity \citep{chen:2019}, a 1-D photochemistry model coupled to a GCM \citep{afrinbadhan:2019,fauchez:2019}, or simply an estimate of the ocean-loss timescale based on the model top water vapor mixing ratios from solely the GCM climate outputs \citep{kopparapu:2016,wolf&toon:2015}. All of these studies estimate the water-loss rates assuming diffusion-limited escape rates of hydrogen \citep{hunten:1973}. The general conclusion seems to be that around inactive stars, the water-loss is more resilient to hydrogen escape, with ocean survival times exceeding $\sim$10~Gyr \citep{Bolmont:2017}. In addition, we have run our own calculations to estimate the water loss and consequent hydrogen escape time scales for all of our aquaplanet cases using a diffusion limited escape rate equation \citep{hunten:1973, wolf&toon:2014}. Our escape time-scale estimates indicate that water loss occurs on extremely long timescales (i.e. tens of billions to a trillion years), consistent with a planet having low water-vapor mixing ratios compared to planets undergoing either a moist or runaway greenhouse phase. These estimates are similar to the ocean lifetime estimates shown in Figure 11 in \citet{kopparapu:2016}, and are consistent with the stellar type dependence shown in that figure. We therefore conclude that TOI-700~d most likely falls within the class of terrestrial planets likely to host high-molecular-weight secondary atmospheres, and a water-rich atmosphere or surface could exist if the initial water inventory for the planet was high.

\subsection{Prospects for Atmospheric Characterization}
\label{sec:prospects}
As one of the nearest M stars with a known transiting Earth-sized planet in the Habitable Zone, TOI-700 is a potential target for future observing campaigns dedicated to characterizing the atmospheres of potentially habitable planets. However, with a stellar radius of 0.41~\rsun, the projected area of TOI-700 is still relatively large compared with more favorable late-type M-star targets; TRAPPIST-1, for example, has a projected size that is 11 times smaller. The transit depth depends linearly on the stellar projected area, and therefore the depth of spectral features in transmission are small - the simulated transmission spectra for all the modeled atmospheres in our study generally have spectral line contrasts that peak at $\sim$6~ppm, a stark illustration of the challenge of characterizing habitable planets even for M-type stars. Dry planets with thin atmospheres have the largest amplitude phase curves, due to a lack of efficient heat transport, both in peak flux as well as amplitude variation, but even these have maximum planet/star flux contrasts of at most $\sim$10~ppm. 

The upcoming launch of the James Webb Space Telescope provides us with our first opportunity to characterize transiting planets across a wide range of wavelengths using a stable space-based platform. However, even the most optimistic scenarios for the noise floor (i.e., the minimum uncertainty on the transit or eclipse depth measurements that can be achieved by JWST due to pseudorandom instrument noise) are 10-20 parts per million (ppm) at 1$\sigma$ \citep{Greene:2016}. Therefore prospects for characterizing TOI-700~d with JWST will most likely be unfeasible, unless the noise floor limit is significantly better than expected. Yet even with a noise floor of $\leq2$~ppm, we would still only be limited to obtaining a constraint on whether the planet has an atmosphere or not \citep{Lustig-Yaeger:2019,koll:2019,fauchez:2019}. Initial estimates of the number of JWST transits or eclipses needed are in the hundreds, and thus would not even fit within the nominal lifetime of the observatory (Rodriguez et al. submitted, and confirmed by our own calculations). In addition, stellar contamination from spots and/or feculae could produce spectral features equal or larger than planetary atmospheric signals at certain wavelengths \citep{ducrot:2018,rackham:2018,zhang:2018}.

Future observatories with ultra-stable instruments and detectors designed specifically for long-baseline time series observations (such as Origins or LUVOIR \citep{origins:2019,luvoir:2019}) may be able to achieve this level of precision, but such performance has not yet been demonstrated; even so, the fundamental challenge of a lack of photons would still exist. However, future planets similar to TOI-700~d may be discovered around stars closer to us. Using PSG, we determine that for an ``Archean Earth'' 1~bar aquaplanet (Case 5), a target similar to TOI-700 would need to be located less than 15~pc away for the CO$_2$ spectral band at \SI{4.3}{\micro\metre} to be detectable in transmission by combining 50 transits, assuming only a photon-noise limit. This hypothetical scenario would require $\sim$320 hours of observing time with an Origins-like telescope.

Similarly, directly imaging TOI-700~d will be prohibitively challenging for even the proposed next-generation high-contrast coronographic instrumentation for the ELTs \citep[such as the PCS/EPICS coronograph proposed for the E-ELT,][]{kasper:2011} or for space-based missions such as the Large Ultraviolet Optical Infrared Surveyor \citep[LUVOIR,][]{roberge:2019}. Imaging the planet from 0.5 $-$ \SI{1}{\micro\metre} would require a contrast of $\sim$10$^{-8}$ at 5~milliarcsec, which is significantly better than the current requirements baselined for these instruments \citep{kasper:2011,roberge:2019}. Significant characterization efforts will therefore require future space-based IR interferometer missions such as the proposed LIFE (Large Interferometer For Exoplanets) mission \citep{quanz:2018}. 

In conclusion, while the detection threshold of the spectral signals for this particular planet are most likely unfeasible for near-term observing opportunities, the end-to-end atmospheric modeling and spectral simulation study that we have performed in this work is an illustrative example of how global climate models can be coupled with a spectral generation model to assess the potential habitability of any HZ terrestrial planets discovered in the future, as we have done here with the exciting new discovery, TOI-700~d. With more discoveries on the horizon with \textit{TESS} and ground-based surveys, we hope that this methodology will prove useful for not only predicting the observability of HZ planets but also for interpreting actual observations in the years to come.

\acknowledgments
The authors acknowledge support from GSFC Sellers Exoplanet Environments Collaboration (SEEC), which is funded in part by the NASA Planetary Science Division’s Internal Scientist Funding Model. E.T.W. also thanks NASA Habitable Worlds grant 80NSSC17K0257. EAG thanks the LSSTC Data Science Fellowship Program, which is funded by LSSTC, NSF Cybertraining Grant \#1829740, the Brinson Foundation, and the Moore Foundation; her participation in the program has benefited this work.

\software{ExoCAM (\url{https://github.com/storyofthewolf/ExoCAM}), PSG \citep{psg:2018}}

\bibliography{bib.bib}

\end{document}